# Analyzing Public Reactions, Perceptions, and Attitudes during the MPox Outbreak: Findings from Topic Modeling of Tweets

Nirmalya Thakur *, Yuvraj Nihal Duggal and Zihui Liu

Department of Computer Science, Emory University, Atlanta, GA 30322, USA;
yuvraj.nihal.duggal@emory.edu (Y.N.D.); chloe.liu2@emory.edu (Z.L.)
* Correspondence: nirmalya.thakur@emory.edu

**Abstract:** In the last decade and a half, the world has experienced outbreaks of a range of viruses such as COVID-19, H1N1, flu, Ebola, Zika virus, Middle East Respiratory Syndrome (MERS), measles, and West Nile virus, just to name a few. During these virus outbreaks, the usage and effectiveness of social media platforms increased significantly, as such platforms served as virtual communities, enabling their users to share and exchange information, news, perspectives, opinions, ideas, and comments related to the outbreaks. Analysis of this Big Data of conversations related to virus outbreaks using concepts of Natural Language Processing such as Topic Modeling has attracted the attention of researchers from different disciplines such as Healthcare, Epidemiology, Data Science, Medicine, and Computer Science. The recent outbreak of the MPox virus has resulted in a tremendous increase in the usage of Twitter. Prior works in this area of research have primarily focused on the sentiment analysis and content analysis of these Tweets, and the few works that have focused on topic modeling have multiple limitations. This paper aims to address this research gap and makes two scientific contributions to this field. First, it presents the results of performing Topic Modeling on 601,432 Tweets about the 2022 Mpox outbreak that were posted on Twitter between 7 May 2022 and 3 March 2023. The results indicate that the conversations on Twitter related to Mpox during this time range may be broadly categorized into four distinct themes—*Views and Perspectives about Mpox*, *Updates on Cases and Investigations about Mpox*, *Mpox and the LGBTQIA+ Community*, and *Mpox and COVID-19*. Second, the paper presents the findings from the analysis of these Tweets. The results show that the theme that was most popular on Twitter (in terms of the number of Tweets posted) during this time range was *Views and Perspectives about Mpox*. This was followed by the theme of *Mpox and the LGBTQIA+ Community*, which was followed by the themes of *Mpox and COVID-19* and *Updates on Cases and Investigations about Mpox*, respectively. Finally, a comparison with related studies in this area of research is also presented to highlight the novelty and significance of this research work.

## 1. Introduction

Monkeypox (Mpox), caused by the monkeypox virus, which belongs to the Poxviridae family, Chordopoxvirinae subfamily, and Orthopoxvirus genus [1], is a re-emerging zoonotic disease. In the shape of a brick-like virion ranging from 200 nm to 250 nm in length, the Mpox virus has a large genome of about 200 kilobase pairs encoding approximately 190 proteins. Two clades of Mpox, clade 1 and clade 2, show a 0.5% genomic difference, with clade 1 having a 1–12% case fatality rate and clade 2 having a 0.1% case fatality rate [2,3]. The first case of human Mpox was recorded in a 9-month-old boy in the Democratic Republic of the Congo (DRC) in 1970 [1]. After the first case in 1970, 59 cases were reported in West and Central Africa in the next decade [4], with a 17% mortality rate in children under 10 [5,6]. The World Health Organization (WHO) monitored Mpox cases

post-1980. Between 1981 and 2017, there were multiple outbreaks of Mpox in DRC due to clade 1, with the fatality rate being between 1 and 12%, primarily due to inadequate health systems [7]. From 2003 to 2022, a few travel-related cases were reported outside endemic countries but the number of cases was not very high [8]. However, a global outbreak of the Mpox virus started on 7 May 2022 [9], and on 23 July 2022, the WHO declared Mpox a Global Public Health Emergency (GPHE) [10]. This outbreak is linked to a new lineage, B.1 (clade 2b), that has a higher mutation rate. This outbreak has resulted in 110 countries reporting about 87,000 cases and 112 deaths so far [11].

The Mpox virus can enter hosts via respiratory or dermal routes. As a result, infection may occur in airway epithelial cells, keratinocytes, fibroblasts, and endothelial cells [12,13]. Currently, no FDA-approved treatments for Mpox exist. At present, in the United States, there are three vaccines—JYNNEOS, ACAM2000®, and APSV—that are available. Out of these three vaccines, JYNNEOS has been approved by the FDA for smallpox and monkeypox in adults at high risk [14]. Tecovirimat is effective against Orthopoxvirus in animals but untested in human Mpox [15]. Other potential treatments for Mpox include VIGIV, cidofovir, and brincidofovir, which have proven in vitro and animal efficacy but limited availability [16]. Since the first case of this outbreak, various policy-making bodies of the world have taken measures to contain the spread of the Mpox virus. For instance, New York City Health + Hospitals (NYC H + H), an integrated healthcare system, has been pivotal in the fight against emerging pathogens and viruses, including Mpox, in the New York City region of the United States [17]. However, the uncertainty and challenges surrounding asymptomatic transmission of viruses such as Mpox and the absence or inadequacy of appropriate and effective transmission-based personal protective equipment (PPE), such as N95 masks, face shields, gowns, and extended cuff examination gloves, may result is a possible resurgence of Mpox [18]. The fifth meeting of the WHO's International Health Regulations (IHR) Emergency Committee on the Multi-Country Outbreak of Mpox took place on 10 May 2023. At this meeting, the committee *acknowledged remaining uncertainties about the disease, regarding modes of transmission in some countries, poor quality of some reported data, and continued lack of effective countermeasures in the African countries, where Mpox occurs regularly*. On 15 May 2023, the US Centers for Disease Control and Prevention (CDC) warned about the potential resurgence of Mpox cases in the US [19].

In today's Internet of Everything era, social media platforms provide a seamless and virtual means for users to connect, communicate, and collaborate with each other [20]. Among the numerous social media platforms that have been used by the public in the last decade and a half, Twitter has been highly popular amongst all age groups. Twitter has over 368 million active monthly users [21]. Twitter stands out as the social media site that journalists choose to use [22] and is among the sites with the highest global adoption rates [23]. On average, 500 million Tweets are published on Twitter each day [24]. Twitter is also the most popular social media platform for news and current events, and the usage of Twitter by Generation Z users is growing approximately 30% faster than for Instagram [25]. Furthermore, on average, users on Twitter who post more than five Tweets follow 405 accounts [26]. Therefore, Twitter has been highly popular amongst researchers from different disciplines for the investigation of a wide range of research problems. During the virus outbreaks that have taken place in the last few years, such as COVID-19, H1N1, flu, Ebola, Zika virus, Middle East Respiratory Syndrome (MERS), measles, and West Nile virus, just to name a few [27], topic modeling of Tweets helped to understand the perception, preparedness, response, views, and opinions of the general public during these virus outbreaks. In a generic manner, topic modeling is a methodology that comprises different algorithms that identify, comprehend, and annotate the thematic structure in a collection of documents [28]. In addition to investigating the perception, preparedness, response, views, and opinions of the general public during these virus outbreaks, the field of topic modeling has had multiple interdisciplinary applications in the last few years, as can be seen from the recent works in topic modeling in bioinformatics [29,30], software engineering [31,32], cryptocurrency [33,34], smart-home research [35,36], human behavior modeling [37,38], analysis of cognitive impairment [39,40], education research [41,42], biology [43,44], and medicine [45,46]. Prior works related to the mining and analysis of Tweets about the MPox outbreak have primarily focused on the sentiment analysis and content analysis of Tweets. Since the first case of this outbreak, only a couple of studies have been published that have focused on topic modeling of the Tweets. However, these studies have multiple limitations centered around (1) the limited time range of the Tweets that were analyzed, (2) the limited number of Tweets that were analyzed, (3) the elimination of a topic from the study, and (4) the lack of reporting of metrics to discuss the working or the accuracy of the topic modeling approaches. Addressing this research gap serves as the main motivation for this work. The rest of the paper is organized as follows. A comprehensive review of recent advances in this area of research is presented in Section 2. Section 3 describes the methodology that was followed for this work. It is followed by Section 4, which presents the results and highlights the novel findings of this work. Section 4 is followed by a conclusion section that summarizes the contributions of this paper and outlines the scope for future work in this area.

## 2. Literature Review

Mining and analysis of Tweets for the investigation and exploration of different research questions, with a specific focus on information diffusion [47], sentiment analysis [48,49], text categorization [50], spam detection [51], privacy issues [52], trust issues [53], user migration [54], and topic detection [55], has been of significant interest to researchers from a wide range of disciplines in the last few years. While misinformation presents some challenges [56], it is still crucial to understand web behavior on Twitter and its implications for real-world decision making. Therefore, this section is section is divided into three parts. Section 2.1 presents a brief review of mining and analysis of Tweets for interdisciplinary research. Section 2.2 discusses the recent advances related to the study and analysis of Tweets in the field of healthcare. Section 2.3 outlines the latest works that have focused on the mining and analysis of Tweets about the MPox outbreak.

*2.1. A Brief Review of Recent Works Related to the Mining and Analysis of Tweets for Interdisciplinary Research*

Abu Samah et al. [57] proposed a web-based dashboard to visualize customer sentiment towards Malaysian airline companies, as expressed on Twitter. Similar to the tourism industry, the entertainment industry and politics have also benefitted from the mining and analysis of Tweets. Bodaghi et al. [58] explored the differences between the web behaviors of actors who spread fake news and those who spread the truth on Twitter. This study showed that although fake news has much better modularity and intra- to interlinks ratio, truth tweeters generally have higher page rank centrality. Collins et al. [59] conducted a comparative analysis of over 2000 tweets from the first two U.S. presidents of the "Twitter era", Barack Obama and Donald Trump, to assess the impact of their online correspondence on America's image abroad and its soft power. An interesting finding of this study was that the tone of presidential tweets can have significant and divergent effects on the perception of the U.S., even internationally. Beyond the border of the United States, Berrocal-Gonzalo et al. [60] found that *politainment*, the phenomenon of trivializing political information for entertainment purposes, manifested on Twitter during the Spanish general elections in April 2019, which prevented the creation of meaningful debates or interactions surrounding the elections on the platform.

The Big Data of conversations on Twitter is also considerably informative regarding the unremitting controversies underlying human society and human rights. Following the United States Supreme Court's decision in Dobbs vs. Jackson Women's Health Organization that overturned abortion rights, Chang et al. [61] presented a large-scale Twitter dataset collected on the abortion rights debate in the U.S. Similarly, Peña-Fernández et al. [62] analyzed the polarization produced in social media debates regarding the rights of transgender people and the views of feminists, specifically the use of the term "TERF" (trans-exclusionary radical feminist) on Twitter. The findings of this work show that online debates are poorly inclusive, suggesting the prevalence of community isolation. Goetz et al. [63] analyzed sentiments in food-security-related Tweets in the U.S. during the early stages of the COVID-19 pandemic, from which they found that keywords of negative emotions were statistically correlated with the contemporaneous food insufficiency rates reported in the Household Pulse Survey. Tao et al. [64] conducted a comparative study of posts on Twitter and Weibo regarding the Russian–Ukrainian War to reveal the differences in the topics of posts between the two platforms and to call for humanitarianism and peace. Researchers have also utilized Twitter as well as studied Tweets in the context of various needs and challenges faced by different diversity groups. In the context of this focus area of research, the needs faced by the elderly population such as falls [65,66], indoor localization issues [67,68], behavior-related problems [69,70], memory issues [71,72], inability to perform activities of daily living (ADLs) [73,74], and user experience with different technologies [75,76], has been widely investigated. Yavuz et al. [65] developed a system for detecting falls, which also comprised virtual support. The virtual support feature included a warning about the fall and specific location-related information of the person who experienced the fall being communicated to caregivers via Twitter messages. Tamplain et al. [69] developed a methodology that could analyze Tweets and identify self-reportings of dyspraxia. An investigation of the evolution and changes in human behavior related to natural disasters in Japan, as expressed on Twitter, was conducted by Lu et al. [73]. Finally, analysis of Tweets has revealed several novel insights underlining the public discourse related to the education industry [77,78], alcohol industry [79,80], tobacco industry [81,82], sports industry [83,84], assisted living technologies [85,86], context-driven technologies [87,88], web behavior analysis [89,90], and emerging works in robotics [91,92]. As can be seen from this brief review, mining and analysis of Tweets holds the potential for the investigation of research questions across different disciplines. In the context of the different virus outbreaks that the world has witnessed in the last decade and a half, healthcare-based research using Tweets has emerged as a crucial utilization of this vast potential of mining and analyzing Tweets. Some recent works are briefly reviewed in Section 2.2, which is followed by a dedicated review of recent works related to the analysis of Tweets about the MPox outbreak.

*2.2. A Brief Review of Recent Works Related to the Mining and Analysis of Tweets for Healthcare Research*

The Big Data of conversations and information exchange from social media platforms, specifically Twitter, has the potential to improve the efficiency, accuracy, and coverage of healthcare systems in different geographic regions. Cevik et al. [93] analyzed the sentiments of Tweets about Parkinson's disease. Kesler et al. [94] applied topic modeling and qualitative content analysis to comments related to cancer-related cognitive impairment (CRCI), revealing the importance of coping mechanisms. Klein et al. [95] demonstrated the potential of using Twitter data to identify the start and end of the 40-week prenatal period, making it a valuable resource for observational studies on potential risk factors in pregnancy. Thakur et al. [96,97] developed a framework to address loneliness and social isolation in the elderly using Twitter data. Thackeray et al. [98] analyzed how Twitter is used during Breast Cancer Awareness Month (BCAM). The findings showed that although organizations and celebrities emphasized fundraisers, early detection, and diagnoses, the general public published the majority of the tweets that did not promote any specific preventive behavior.

Studies in this field have shown that Tweets provide a timely and authentic record of public perception and understanding of human health crises. To investigate how the Dengue epidemic was reflected on Twitter, Gomide et al. [99] proposed an active surveillance methodology based on volume, location, time, and public perception. The analysis found a high correlation between the number of cases reported by official statistics and the number of Tweets posted during the same period. The work performed by Radzikowski et al. [100] reported that news organizations had a higher impact than health organizations in communicating health-related information. By examining the use of Twitter data to track public sentiment and disease activity related to H1N1 or swine flu, Signorini et al. [101] found that estimates of influenza-like illness derived from Twitter chatter accurately tracked reported disease levels from governmental organizations. With the realization of Twitter's ability to provide real-time information, Sugumaran et al. [102] collected and analyzed tweets tagged with #WestNileVirus and #WNV and concluded that unusually higher temperatures and mosquito activities led to an increase in Tweet numbers about the West Nile virus. To inform health promotion efforts, Porat et al. [103] analyzed the content and sources of popular Tweets related to a diphtheria case in Spain. The most notable conclusion from their study was their suggestion for healthcare organizations to collaborate with popular journalists, news outlets, and science authors to address public concerns and misinformation through the outlet of social media platforms such as Twitter. As can be seen from this brief review, the study, analysis, and interpretation of multi-modal characteristics of Tweets about different virus outbreaks has helped in the timely advancement of research in the field of healthcare. Section 2.3 specifically highlights the recent advances in this area of research that focused on the investigation of the public discourse on Twitter about MPox.

*2.3. Review of Recent Works Related to the Mining and Analysis of Tweets about MPox*

Knudsen et al. [104] studied 262 Tweets to describe the risk of Mpox to students. The results showed that credentialed Twitter users were 4.6 times more likely to Tweet inaccurate information about MPox. Zuhanda et al. [105] studied 5000 Tweets about MPox posted on 5 August 2022 to perform sentiment analysis. The results showed that 51.92% of the Tweets had a negative sentiment and 48.08% of the Tweets had a positive sentiment. Ortiz-Martínez et al. [106] performed a study of top 100 Tweets about MPox posted on 24 May 2022. The findings showed that most of the Tweets were posted by informal individuals or groups (60%), followed by healthcare or public health (32%), and news outlets or journalists (8%). The work by Rahmanian et al. [107] involved studying 384,560 Tweets posted between 16 May 2022 and 22 May 2022. The results indicated that most of these Tweets were posted by individuals from the United States and Canada. Cooper et al. [108] studied Tweets containing the word "monkeypox" posted between 1 May 2022 and 23 July 2022. The results showed that a total of 48,330 Tweets were posted by individuals who identified as members or allies of the LGBTQ+ community. The work of Ng et al. [109] focused on studying 352,182 Tweets about MPox posted between 6 May 2022 and 23 July 2022. The authors performed topic modeling of these Tweets and derived three themes—concerns of safety, stigmatization of minority communities, and a general lack of faith in public institutions. Bengesi et al. [110] mined over 500,000 multilingual tweets related to MPox and performed sentiment analysis. Olusegun et al. [111] studied 800,000 Tweets about MPox and used NRCLexicon to predict and measure the emotional significance of each Tweet. Farahat et al. [112] performed opinion mining on a total of 8532 Tweets about MPox posted between 22 May 2022 and 5 August 2022. The results indicated that close to 50% of the Tweets were neutral. Sv et al. [113] studied Tweets about MPox posted between 1 June 2022 and 25 June 2022. The study was performed in two steps. In the first step, the authors analyzed a dataset of 556,402 Tweets about MPox and performed sentiment analysis of those Tweets. The results showed that 128,037 Tweets (23.01%) had a negative sentiment. Thereafter, only these 128,037 Tweets that had a negative sentiment were used for topic modeling in the second step of their study. The results of topic modeling showed that among the Tweets about MPox that had negative sentiments, there was a range of topics that were represented, such as deaths caused by the MPox virus, the severity of the virus, lesions caused by the virus, whether the virus is airborne, vaccines for the virus, and whether the virus will lead to the next pandemic. Mohbey et al. [114] developed a neural-network-based model to perform opinion mining of Tweets about MPox, and the accuracy of the model was found to be 94%. A dataset of Tweets about the MPox outbreak was developed by Nia et al. [115]. The work of Iparraguirre-Villanueva [116] focused on the detection of polarity in conversations on Twitter about MPox. The results showed that 45.42% of people expressed neither positive nor negative opinions, whereas 19.45% expressed negative and fearful feelings about MPox. AL-Ahdal [117] studied 15,936 Tweets about MPox posted by individuals from Germany. The results showed that the public displayed an impersonal feeling toward MPox.

As can be seen from this review and the research questions that were investigated in these papers, the major focus areas of these works have been sentiment analysis and content analysis. Only a couple of works have focused on the topic modeling of Tweets about MPox. However, those works have multiple limitations centered around (1) the limited time range of the Tweets that were analyzed, (2) the limited number of Tweets that were analyzed, (3) the elimination of a topic from the study, and (4) the lack of reporting of metrics to discuss the working or the accuracy of the topic modeling approaches (discussed in detail in Section 4). This study aims to address this research gap. Therefore, topic modeling of 601,432 Tweets about the 2022 MPox outbreak that were posted on Twitter between 7 May 2022 and 3 March 2023 was performed in this study. Section 3 outlines the step-by-step process that was followed for the system design and implementation. The results and novel contributions of this work are presented and discussed in Section 4.

## 3. Methodology

This section presents the methodology that was followed for the system design and implementation. This section is divided into three parts. In Section 3.1, a technical overview of RapidMiner [118] is presented, as RapidMiner was used for this work. Section 3.2 presents the description of the topic modeling architecture that was used in this work. Section 3.3 outlines the steps that were followed for the implementation of this topic modeling architecture, along with the specifics of the system design.

### 3.1. Technical Overview of RapidMiner

RapidMiner is a Data Science software platform that allows the development and implementation of different algorithms related to Machine Learning, Data Science, Artificial Intelligence, and Big Data. It enables its users to visually design data workflows and build predictive models using a graphical user interface (GUI). Applications developed in RapidMiner are referred to as "processes" that comprise one or more "operators". These "operators" may be built-in or user defined. The RapidMiner Studio consists of several built-in "operators" that may be directly used for the implementation of various tasks. There also exist certain "operators" in RapidMiner Studio that may be utilized to modify the functionality of other "operators". RapidMiner is developed on a client–server model with public and private cloud infrastructures. There is a free version and an enterprise version of RapidMiner Studio. The free version has a limit of 10,000 rows in a dataset for a "process". For this research work, the educational license of RapidMiner (available to researchers in academia upon request) was used in RapidMiner Studio 10.1.001. With the educational license, RapidMiner Studio can be used to process any number of rows in any dataset. The following represent a few notable characteristics of RapidMiner Studio [118,119]:

1. It supplies pre-built "operators" encompassing distinct functions that can be directly employed or customized for the creation and execution of algorithms and applications.
2. RapidMiner is developed using Java, which ensures that RapidMiner "workflows" retain the write once run anywhere (WORA) attribute of Java.
3. The platform permits the installation of various extensions to facilitate seamless connectivity and integration of RapidMiner "workflows" with other software and hardware environments.

4.  Scripts developed in programming languages, such as Python and R, can also be imported into a RapidMiner "workflow" to supplement its functionalities.
5.  The software enables the creation of new "operators" and effortless dissemination of the same within the RapidMiner community.
6.  RapidMiner consists of "operators" that enable it to establish connections with social media platforms, such as Twitter and Facebook. Such connections facilitate the extraction of tweets, comments, posts, reactions, and other relevant social media interactions.

Similar to RapidMiner, there are other data science platforms, for example, WEKA [120] and MLC++ [121]. Both these platforms can be utilized for developing machine learning models. However, a major limitation of these platforms is that nesting of "operators" is not permitted. Such a limitation is not present in RapidMiner. In view of this and in view of the different features of RapidMiner related to the development and implementation of machine learning algorithms and models, RapidMiner was utilized for this research work.

*3.2. Description of the Topic Modeling Architecture for System Design*

Latent Dirichlet Allocation (LDA) [122] is a generative probabilistic model used in the field of Natural Language Processing and Machine Learning. It is commonly used for topic modeling, which is the task of identifying topics within a collection of documents. In LDA, the mixture of topics is derived from a consistent Dirichlet prior, which is the same across all documents. The procedure [123] for creating a corpus is outlined as follows (in this context, a smoothed LDA is considered). Thereafter, the likelihood of generating a corpus can be represented, as shown in Equation (1).

1.  Select a multinomial distribution $\phi_z$ for each topic z from a Dirichlet distribution with parameter $\beta$.
2.  For every document d, select a multinomial distribution $\theta_d$ from a Dirichlet distribution with parameter $\alpha$.
3.  In document d, for each word w, select a topic z, such that z ∈ {1….K} from the multinomial distribution $\theta_d$.
4.  Select w from the multinomial distribution $\theta_z$.

$$P(Doc_1,\ldots,Doc_N|\alpha,\beta) = \int\int \prod_{z=1}^{K} P(\phi_z|\beta) \prod_{d=1}^{N} P(\theta_d|\alpha) \left(\prod_{i=1}^{N_d}\sum_{z_i=1}^{K} P(z_i|\theta)P(w_i|z,\phi)\right) d\theta d\phi \quad (1)$$

LDA considers the topic distribution as a k-parameter hidden random variable instead of a large set of features associated with the training data to address the problems of overfitting and generating new documents that were present in probabilistic latent semantic indexing (pLSI) [123]. To use language models for information retrieval in an LDA, an approach using the query likelihood model is used, where each document is scored by the likelihood of its model generating a query Q. This is shown in Equations (2) and (3). In Equation (2), D represents a model for documents, Q stands for the query, and q denotes an individual term within the query Q. P(Q|D) represents the probability of the document model generating the query terms, following the assumption of "bag-of-words". This assumption considers that terms are independent. $P(q_i|D)$ is specified by the document model with Dirichlet smoothing. In Equation (3), P(w|D) is the maximum likelihood estimate of word w in the document D, P'(w|coll) is the maximum likelihood estimate of the same word w in the entire collection, and µ represents the Dirichlet prior. It is important to mention here that each topic in an LDA model represents a specific combination of words. However, the same approach may not always be as accurate as the working of non-topic models in Natural Language Processing such as unigram or bigram analysis. As a result, directly implementing the LDA model may hurt the overall performance of information retrieval. So, in a prior work in this field, the original document model (Equation (3)) was combined with the LDA model to construct a new LDA-based document model, as shown in Equation (4). The LDA model introduces a novel document representation centered around topics. After obtaining the posterior estimates for θ and φ, the word probability within a document can be computed using Equation (5), where $\hat{\theta}$ and $\hat{\phi}$ are the posterior estimates of $\theta$ and $\emptyset$, respectively [123].

$$P(Q|D) = \prod_{q \in Q} P(q|D) \tag{2}$$

$$P(w|D) = \frac{N_d}{N_d + \mu} P_{ML}(w|D) + \left(1 - \frac{N_d}{N_d + \mu}\right) P_{ML}(w|coll) \tag{3}$$

$$P(w|D) = \lambda \left(\frac{N_d}{N_d + \mu} P_{ML}(w|D) + \left(1 - \frac{N_d}{N_d + \mu}\right) P_{ML}(w|coll)\right) + (1 - \lambda) P_{lda}(w|D) \tag{4}$$

$$P_{lda}(w|d, \hat{\theta}, \hat{\phi}) = \sum_{z=1}^{K} P(w|z, \hat{\phi}) P(z|\hat{\theta}, d) \tag{5}$$

The LDA cannot be solved by direct inference. So, Gibbs sampling is utilized, which helps in the approximation of $\hat{\theta}$ and $\hat{\phi}$, with α and β being the hyperparameters that determine the smoothness of the empirical distribution. Gibbs sampling involves performing an iteration over a set of variables $z_1, z_2, z_3, \ldots z_n$, where, for each iteration, $z_i$ is sampled from $P(z_i|z_{\setminus i}, w)$. Every such iteration over all these variables is known as a Gibbs sweep. After a considerable number of iterations, the Gibbs sampling for an LDA produces samples from $P(z|w)$. This sampling may be performed by jointly resampling all the topics. In this approach, the scope of a Gibbs sweep is defined to be the hidden topic variables by taking into account both original and new documents. Initially, the sampling of the topic variables for the training set is performed such that they converge (without the new documents). Thereafter, the topic variables are randomly initialized and the sampling is performed again such that the model converges by taking into account all the documents. At this point, the topic distribution, $\theta_d$ can be estimated using a single Markov chain state, as shown in Equation (6). In this Equation, $n_{.|d}$ represents the length of the document.

$$(\hat{\theta}_{t|d}) = \frac{\alpha_t + n_{t|d}}{\sum_{t'} \alpha_{t'} + n_{.|d}} \tag{6}$$

A higher accuracy may be obtained by computing the average of values generated by Equation (6) from multiple Markov chains [123]. In this research work, SparseLDA was implemented, as prior work has shown that it is 20 times faster than the traditional LDA [124]. In the SparseLDA framework, given an observed word type w, the probability of topic z in document d can be computed using Equation (7).

$$P(z = t|w) \alpha \left(\alpha_t + n_{(t|d)}\right) \frac{\beta + n_{w|t}}{\beta V + n_{.|t}} \tag{7}$$

In this context, the process of sampling involves the calculation of the unnormalized weight, q(z), for each topic by sampling a random variable $U \sim U(0, \sum_z q(z))$ and evaluating t such that $\sum_{z=1}^{t-1} q(z) < U < \sum_{z=1}^{t} q(z)$. The procedure also necessitates the calculation of q(z) for all the topics for the computation of the normalizing constant for the distribution $\sum_z q(z)$, despite the fact that the probability mass is usually concentrated on a small set of topics. A simpler approach involves caching a significant portion of the computation needed to calculate the normalizing constant. Through the reorganization of terms within the numerator, Equation (7) can be portioned into three distinct sections, as shown in Equations (9)–(11). Here, the first term is constant for all documents and the second term does not depend on the current word type. Moreover, $\sum_z q(z)$ corresponds to the sum across topics for each of the three components in Equation (8) [124].

$$P(z = t|w) \alpha \frac{\alpha_t \beta}{\beta V + n_{.|t}} + \frac{n_{t|d} \beta}{\beta V + n_{.|t}} + \frac{(\alpha_t + n_{t|d}) n_{w|t}}{\beta V + n_{.|t}} \tag{8}$$

$$s = \sum_t \left( \frac{\alpha_t \beta}{\beta V + n_{.|t}} \right) \tag{9}$$

$$r = \sum_t \left( \frac{n_{t|d} \beta}{\beta V + n_{.|t}} \right) \tag{10}$$

$$q = \sum_t \left( \frac{(\alpha_t + n_{t|d}) n_{w|t}}{\beta V + n_{.|t}} \right) \tag{11}$$

This divides the full sampling mass into three "buckets". Now, $U \sim \mathcal{U}(0, s + r + q)$ can be sampled. If U < s, it would imply hitting the "smoothing-only" bucket. Thereafter, the process involves stepping through each topic and calculating and adding $\frac{\beta \alpha_t}{\beta |V| + n_{.|t}}$ for that topic until it is greater than x. For the document bucket, s < x < (s + r), the process involves iterating through the set of topics that satisfies $n_{t|d} \neq 0$. The scenario of x > (s + r) implies hitting the "topic word" bucket, and topics are considered such that $n_{w|t} \neq 0$. The calculation of the three parameters of the normalizing constant, r, s, and q, is not complicated. The constant s only changes when the hyperparameter $\alpha$ is updated. Conversely, the constant r is solely influenced by document topic counts. This permits the computation of r once at the start of each document and its subsequent modification by subtracting and adding values related to the prior and current topics in each Gibbs update. This process takes constant time and is not dependent on the number of topics. The topic word constant, q, changes with the value of w, so old computations cannot be easily recycled. However, the performance can be significantly improved by splitting q into two components, as shown in Equation (12). With this Equation, the coefficient $\frac{(\alpha_t + n_{t|d})}{\beta V + n_{.|t}}$ can be cached for every topic. Calculating q for a specific w involves performing a single multiplication operation for each topic, where $n_{w|t} \neq 0$. Given that $n_{t|d} = 0$ for all topics within any document, the coefficients vector will predominantly comprise only $\frac{(\alpha_t)}{\beta |V| + n_{.|t}}$. Therefore, this allows the optimization of the LDA model by caching these coefficients across documents, refreshing values only for topics with non-zero counts in the current document, and reverting these values to $\alpha$-only values upon finishing the sampling process for that document.

$$q = \sum_t \left[ \frac{(\alpha_t + n_{t|d})}{\beta V + n_{.|t}} \times n_{w|t} \right] \tag{12}$$

If the values of $\alpha$ and $\beta$ are small, most of the total mass is taken up by q. It has been found empirically that 90% of the samples belong to this "bucket". In a Dirichlet-multinomial distribution comprising fewer parameter magnitudes, the likelihood of the distribution is approximately proportional to the concentration of the counts on a small number of dimensions. It has also been found that the time per iteration is approximately proportional to the likelihood of the model. As the sampler approaches a region of high probability, the time per iteration decreases, leveling off as the sampler converges. The efficiency of this algorithm depends on its capability to detect topics by satisfying the condition of $n_{w|t} \neq 0$. In addition to this, as the expression in Equation (12) is approximately proportional to $n_{w|t}$ and as the evaluation of the terms can be stopped from the point where the sum of the terms is more than U − (s + r), it is considered desirable to perform the iteration over non-zero topics in descending order [124].

*3.3. Description of the System Design and Implementation*

This section describes the system design and its implementation, as well as the dataset that was used for performing this research work. The dataset that was used comprises 601,431 Tweet IDs of Tweets about MPox posted between 7 May 2022 and 3 March 2023 [125]. This dataset contains only Tweet IDs, and the standard procedure for working with such datasets is that the dataset is hydrated to obtain the text of the Tweets and related information. However, this dataset was developed by the first author of this paper, so all the Tweets were already available and hydration of the Tweet IDs was not necessary. Figure 1 shows the system design in RapidMiner. This is a "process" that was developed in RapidMiner Studio 10.1.001 (with the Educational License) to set up this system, and this "process" comprises different "operators" with different functionalities.

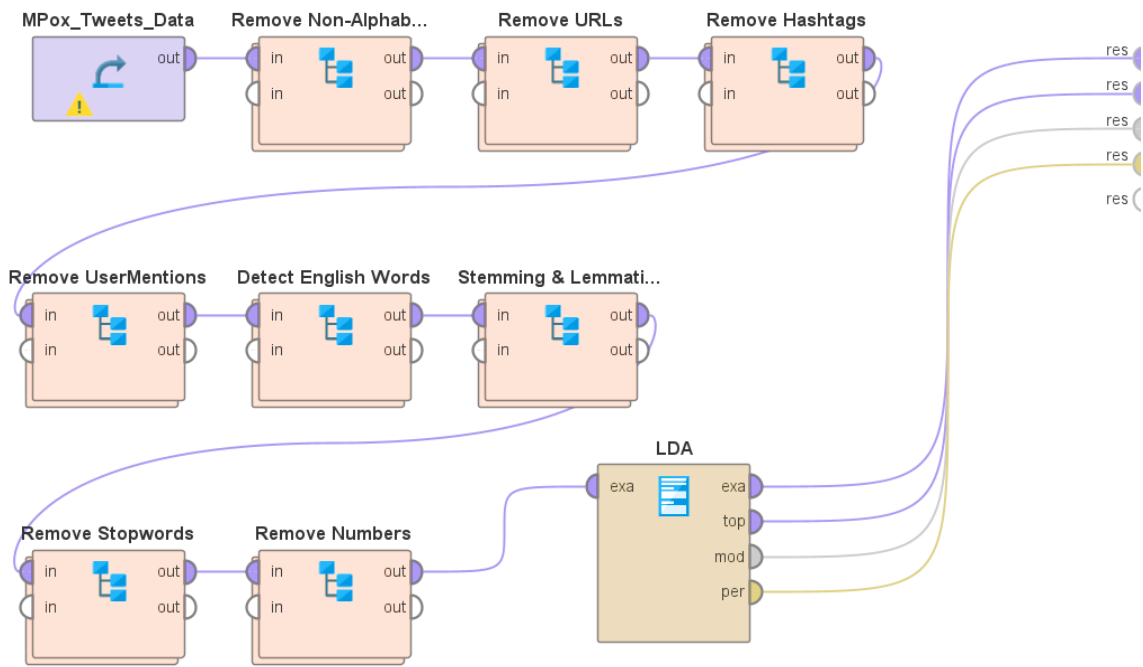

**Figure 1.** System Design in RapidMiner for Performing Topic Modeling.

In this Figure, the "MPox_Tweets_Data" "operator" represents the Tweets from the dataset described earlier in this section. These Tweets were imported into RapidMiner Studio to develop this "process". All 601,431 Tweets about MPox posted between 7 May 2022 and 3 March 2023 were used for developing this LDA model. Thereafter, separate "operators" were developed to perform the different steps of the data processing. The data processing comprised the following steps, and for each of these steps, a separate "operator" was developed in this RapidMiner "process". For steps (a), (b), (c), and (d) of the data preprocessing, different regular expressions (RegEx) were developed and applied to define the functionalities of these "operators".

(a) Removal of characters that are not alphabets (RegEx used: ^a-zA-Z).
(b) Removal of URLs (RegEx used: http\S+).
(c) Removal of hashtags (RegEx used: #[A-Za-z0-9]).
(d) Removal of user mentions (RegEx used: @[A-Za-z0-9]).
(e) Detection of English words using tokenization.
(f) Stemming and Lemmatization.
(g) Removal of stop words.
(h) Removal of numbers.

After completion of the data pre-processing, an LDA model was developed and implemented in RapidMiner as per the architecture of the parallel topic model and SparseLDA (described in Section 3.2) by customizing and utilizing the "Extract Topics from Data (LDA)" operator in RapidMiner Studio 10.1.001. The number of iterations for optimization was set to 1000, as using 1000 iterations for developing LDA models has been used in related studies in this area of research [126,127], where 1000 iterations were suggested to reach a stable convergence. There are three hyperparameters that are associated with the LDA model that was implemented—the number of topics (k), the distribution of words per topic ($\beta$), and the distribution of topics per document ($\alpha$). The distribution of topics per document ($\alpha$) was set to 50/k and the distribution of words per topic ($\beta$) was set to 50/(number of words), which are standard considerations related to topic modeling using LDA [127]. The frequency of hyperparameter optimization was set to 10, which is the default value for this "operator" in RapidMiner Studio 10.1.001. As discussed in related studies in this area of research [128,129], the average coherence value of an LDA model serves as a key indicator for the determination of the optimal number of topics. So, this "process" (shown in Figure 1) was repeatedly run by varying the number of topics from 2 to 50, and the average coherence value of the model was computed and recorded for each of these runs. The results of running this "process" 49 times to deduce the optimal number of topics, as well as the specific topics that were identified in the Tweets, are discussed in Section 4. Figure 2 shows the order in which the different "operators" of this "process" were executed for each run of this "process".

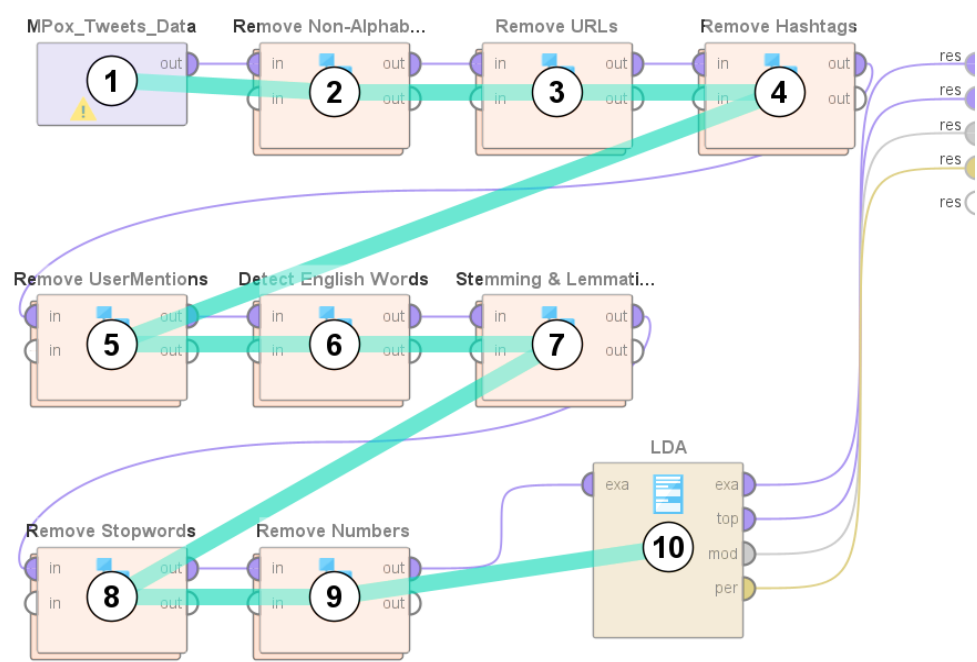

**Figure 2.** Representation of the order of execution of the different "operators" of the developed "process" in RapidMiner.

## 4. Results and Discussions

This section presents the results of this work. As stated in Section 3.3, the LDA model (shown in Figure 1) was run by varying the number of topics from 2 to 50 to determine the optimal number of topics based on the analysis of the average coherence value for each run. Table 1 represents the average coherence value of this LDA model from each run. This table was compiled from 49 runs of this LDA model by varying the number of topics from 2 to 50. An analysis of the same is presented in Figure 3.

**Table 1.** Average coherence values of the LDA model shown in Figure 1 for different numbers of topics.

| Number of Topics | Average Coherence Value |
| --- | --- |
| 2 | −6.1450 |
| 3 | −6.0560 |
| 4 | −4.6730 |
| 5 | −5.2120 |
| 6 | −5.7230 |
| 7 | −5.8700 |
| 8 | −6.6150 |
| 9 | −6.8800 |
| 10 | −6.1840 |
| 11 | −5.6000 |
| 12 | −5.4140 |
| 13 | −5.3280 |
| 14 | −6.7830 |
| 15 | −6.0380 |
| 16 | −5.6930 |
| 17 | −5.6520 |
| 18 | −6.5670 |
| 19 | −6.0470 |
| 20 | −6.0610 |
| 21 | −6.3420 |
| 22 | −5.5790 |
| 23 | −6.0700 |
| 24 | −5.9090 |
| 25 | −6.7030 |
| 26 | −6.6010 |
| 27 | −5.9930 |
| 28 | −5.9870 |
| 29 | −5.8120 |
| 30 | −6.0040 |
| 31 | −5.8810 |
| 32 | −6.0350 |
| 33 | −5.8860 |
| 34 | −6.2010 |
| 35 | −5.7920 |
| 36 | −6.0450 |
| 37 | −6.5680 |
| 38 | −6.3470 |
| 39 | −6.1800 |
| 40 | −6.2180 |
| 41 | −6.4490 |
| 42 | −6.1700 |
| 43 | −6.2120 |
| 44 | −6.3390 |
| 45 | −5.8690 |
| 46 | −6.2330 |
| 47 | −6.1720 |
| 48 | −5.9840 |
| 49 | −6.1210 |
| 50 | −5.9280 |

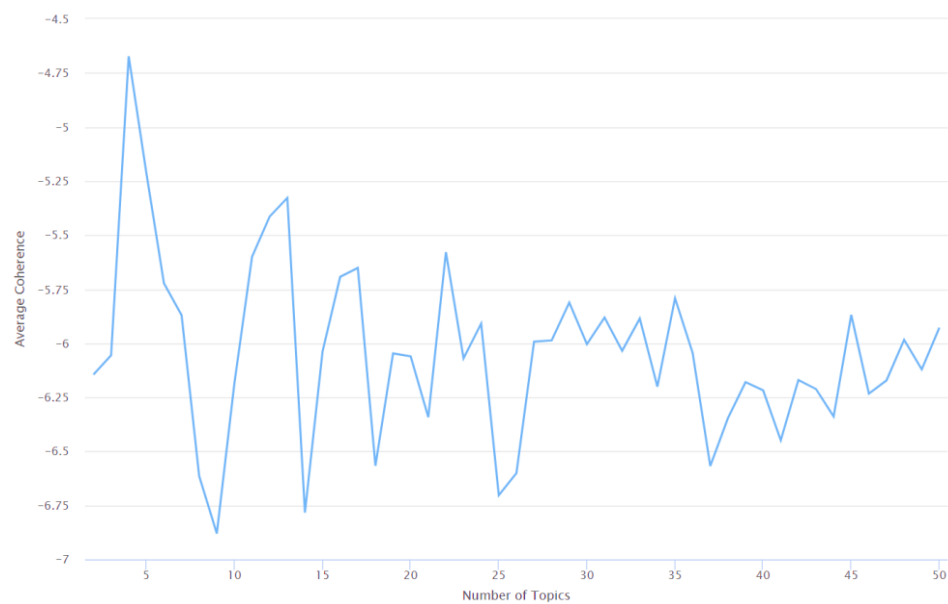

**Figure 3.** Analysis of the average coherence values of the LDA model for different numbers of topics.

From Table 1 and Figure 3, the optimal number of topics was determined to be four, as the LDA model produced the highest coherence score for the same. It is worth mentioning here that negative values of coherence scores are not unusual for an LDA model that follows the system architecture as described in this paper, as the formula for computing the coherence scores (for the LDA model used in this work) involves applying the logarithmic function on calculated probability values, as shown in Equation (13), where C stands for the average coherence score, N represents the top words of a topic, and $w_i$ and $w_j$ represent the *i*th and *j*th word, respectively. In addition to this, $\varepsilon$ features in this Equation to prevent a scenario of the logarithmic function being applied to zero [130].

$$C = \frac{2}{N(N-1)} \sum_{i=2}^{N} \sum_{j=1}^{i-1} \log \frac{P(w_i, w_j) + \varepsilon}{P(w_j)} \quad (13)$$

Furthermore, a recent study [131] followed a similar system architecture for the LDA model that was developed, and the authors of that work obtained negative values for the coherence scores for all the numbers of topics; the lowest value of the coherence score was reported to be as low as about −11.5. After determining that the optimal number of topics for this LDA model was four, this model was run by setting the number of topics as four, and the characteristics from the output were observed and analyzed. For each Tweet, the RapidMiner "process" computed a confidence value for each topic and then predicted a topic for that Tweet based on the highest confidence value. This is shown in Figure 4. To avoid an output table comprising 601,431 rows, this Figure shows 17 rows (selected randomly) from this output table. In Figure 4, the attributes *confidence(Topic_0)*, *confidence(Topic_1)*, *confidence(Topic_2)*, and *confidence(Topic_3)* represent the confidence values of each Tweet belonging to Topics 0, 1, 2, and 3, as computed by the LDA model. The last attribute in this Figure, *Tweet_Text*, shows the Tweet (after data preprocessing) that was analyzed. For each Tweet, the highest of these confidence values was used to predict the topic for the same. For instance, in the first row in Figure 4, it can be seen that *confidence(Topic_0)* has the highest value, so the predicted topic for this Tweet was Topic 0. In a similar manner, the LDA model predicted the topics for all the Tweets of this dataset.

| Row No. | documentid | prediction(Topic) | confidence(Topic_0) | confidence(Topic_1) | confidence(Topic_2) | confidence(Topic_3) | Tweet_Text |
|---|---|---|---|---|---|---|---|
| 117 | 116 | Topic_0 | 0.739 | 0.062 | 0.143 | 0.056 | also been preoccupied wit… |
| 118 | 117 | Topic_1 | 0.049 | 0.674 | 0.028 | 0.250 | Tomorrow on Morning wit… |
| 119 | 118 | Topic_0 | 0.708 | 0.023 | 0.258 | 0.011 | I already done enough mo… |
| 120 | 119 | Topic_0 | 0.852 | 0.018 | 0.090 | 0.040 | The monkey pox they bee… |
| 121 | 120 | Topic_0 | 0.867 | 0.027 | 0.054 | 0.052 | That s what they said abo… |
| 122 | 121 | Topic_0 | 0.685 | 0.013 | 0.064 | 0.239 | people have to take of Co… |
| 123 | 122 | Topic_0 | 0.414 | 0.384 | 0.175 | 0.027 | First it will be that monkey … |
| 124 | 123 | Topic_3 | 0.345 | 0.022 | 0.139 | 0.495 | don t sweat the Monkey Pox |
| 125 | 124 | Topic_0 | 0.739 | 0.018 | 0.084 | 0.160 | All you are getting monkey… |
| 126 | 125 | Topic_0 | 0.798 | 0.008 | 0.036 | 0.158 | This Monkey Pox I d rather… |
| 127 | 126 | Topic_3 | 0.221 | 0.013 | 0.098 | 0.668 | anyone have any Monkey … |
| 128 | 127 | Topic_0 | 0.504 | 0.010 | 0.173 | 0.313 | I have no idea what s goin… |
| 129 | 128 | Topic_0 | 0.908 | 0.015 | 0.041 | 0.037 | if that monkey pox hit the l … |
| 130 | 129 | Topic_2 | 0.473 | 0.005 | 0.491 | 0.032 | Why are u people showin… |
| 131 | 130 | Topic_0 | 0.887 | 0.013 | 0.053 | 0.047 | The same who he already … |
| 132 | 131 | Topic_3 | 0.235 | 0.022 | 0.066 | 0.677 | Bunch of Monkey Pox now |
| 133 | 132 | Topic_0 | 0.602 | 0.013 | 0.043 | 0.342 | MONKEY POX I wonder if t… |
| 134 | 133 | Topic_0 | 0.881 | 0.010 | 0.049 | 0.060 | I it because i was riding th… |

**Figure 4.** A selection of 17 rows (selected randomly) from the output table of the developed LDA model.

It is important to mention that every Tweet in this dataset was assigned a confidence value for every topic (Topic 0, Topic 1, Topic 2, and Topic 3) and then the highest confidence value out of these four confidence values was utilized by the developed LDA model to determine the specific topic for a Tweet. The minimum, maximum, average, and standard deviation of these confidence values for each topic are presented in Table 2. Figure 5 shows a histogram-based representation of the confidence values for Topic 0 for all the analyzed Tweets and the categorization of these Tweets into different topics. The X-axis represents the confidence values, and the Y-axis represents the number of Tweets. The color coding (as stated in this Figure) represents the predicted topic by the LDA model. As can be seen from Figure 5, all the Tweets that received a confidence value of around 0.5 or higher for Topic 0 were categorized as Topic 0. Figures 6–8 show similar histogram-based representations of the confidence values of Tweets for Topics 1, 2, and 3, respectively, and their categorization into different topics.

**Table 2.** The minimum, maximum, average, and standard deviation of the confidence values associated with each topic.

| Topic | Minimum Confidence Value | Maximum Confidence Value | Average Confidence Value | Standard Deviation of the Confidence Value |
|---|---|---|---|---|
| Topic 0 | 0.016 | 0.989 | 0.584 | 0.280 |
| Topic 1 | 0.002 | 0.940 | 0.083 | 0.180 |
| Topic 2 | 0.005 | 0.951 | 0.179 | 0.200 |
| Topic 3 | 0.005 | 0.978 | 0.154 | 0.200 |

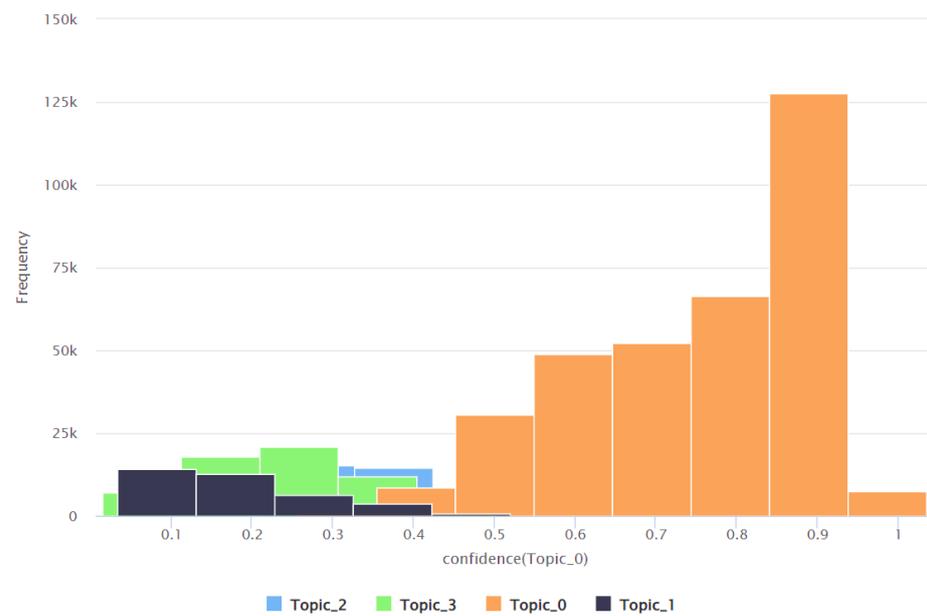

**Figure 5.** A histogram-based representation of the confidence values (for Topic 0) for all the analyzed Tweets and the categorization of these Tweets into different topics (the color coding represents the categorization of these Tweets into different topics).

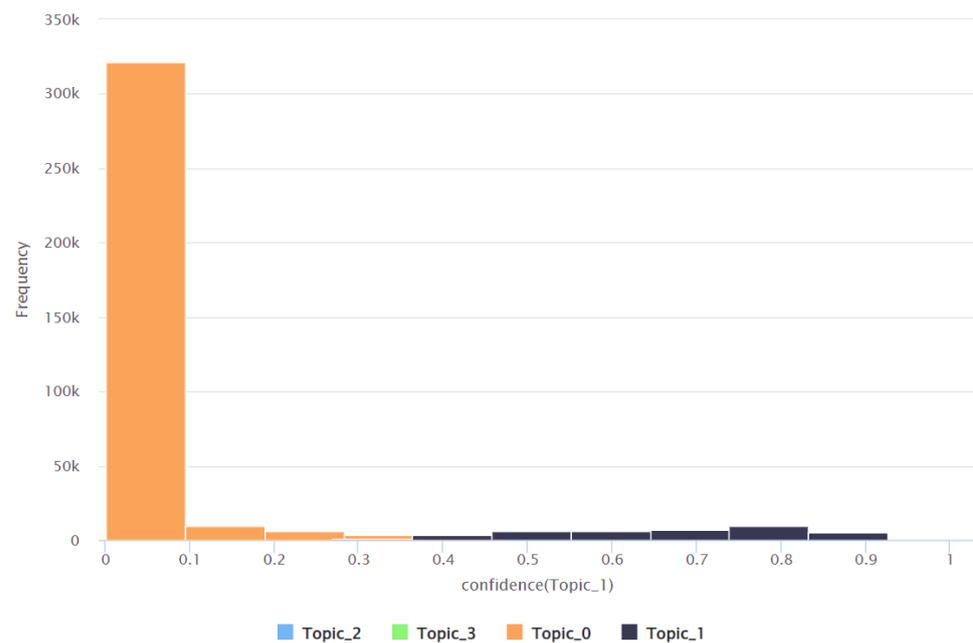

**Figure 6.** A histogram-based representation of the confidence values (for Topic 1) for all the analyzed Tweets and the categorization of these Tweets into different topics (the color coding represents the categorization of these Tweets into different topics).

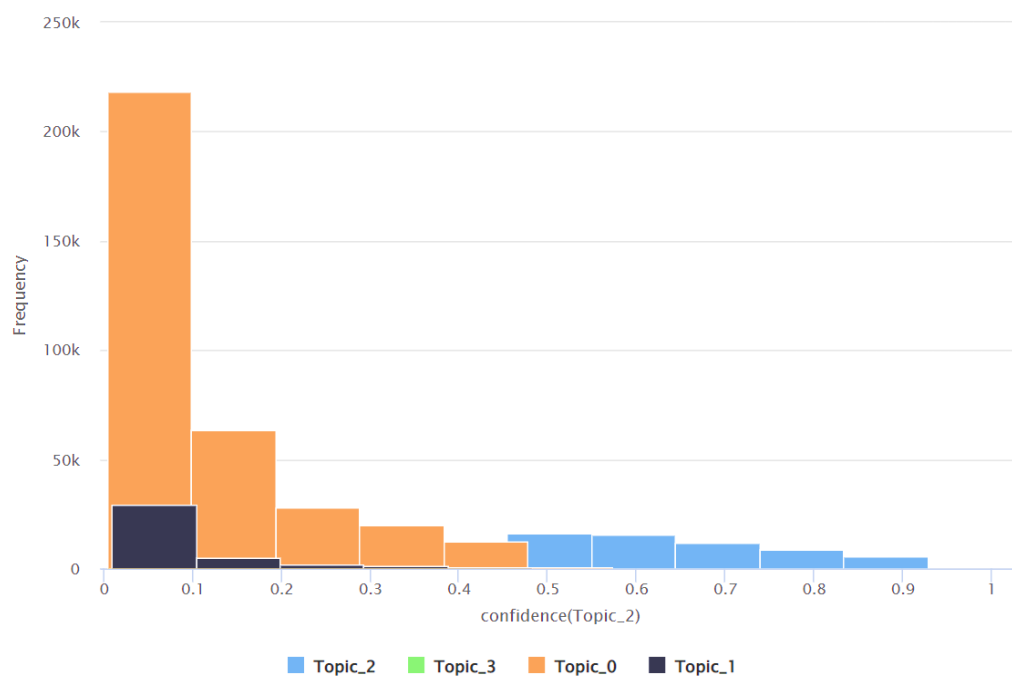

**Figure 7.** A histogram-based representation of the confidence values (for Topic 2) for all the analyzed Tweets and the categorization of these Tweets into different topics (the color coding represents the categorization of these Tweets into different topics).

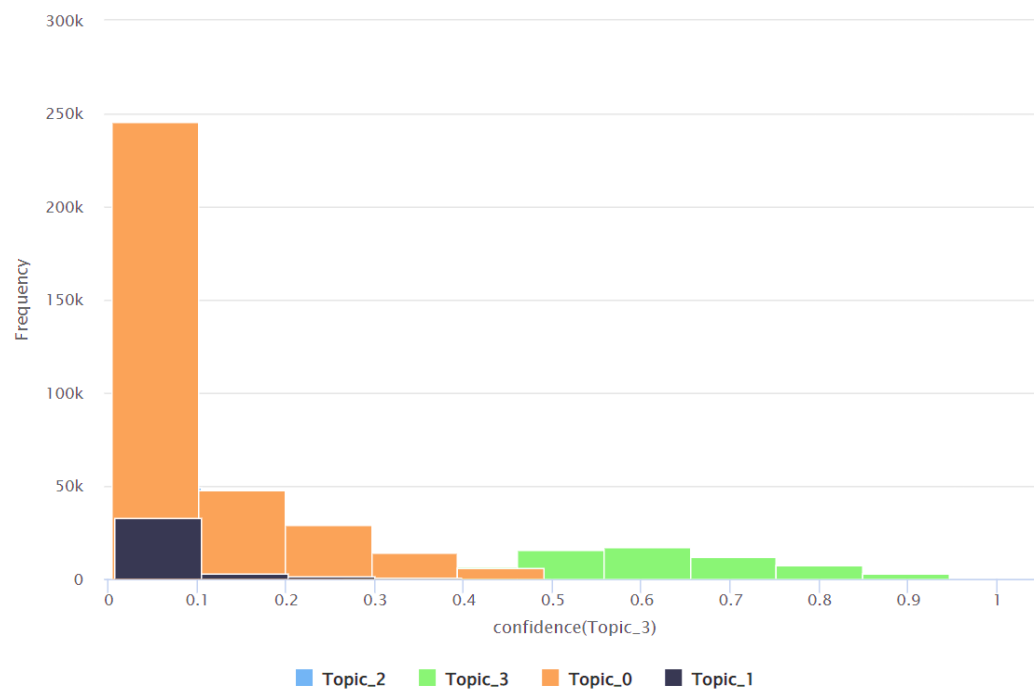

**Figure 8.** A histogram-based representation of the confidence values (for Topic 3) for all the analyzed Tweets and the categorization of these Tweets into different topics (the color coding represents the categorization of these Tweets into different topics).

The Tweets belonging to each of these topics—Topic 0, Topic 1, Topic 2, and Topic 3—were studied to understand the underlying themes of conversation that represented each of these topics. Based on this study, the broad themes that represented these topics were observed to be *"Views and Perspectives about MPox"*, *"Updates on Cases and Investigations about Mpox"*, *"MPox and the LGBTQIA+ Community"*, and *"MPox and COVID-19"*.

Table 3 represents a random selection of five Tweets for each of these Topics. In Table 3, these Tweets are presented in "as is" form, i.e., in the manner in which they were originally posted on Twitter to provide better context.

**Table 3.** Representation of five Tweets (selected randomly) for each Topic—Topic 0, Topic 1, Topic 2, and Topic 3.

| Tweet # | Original Text of the Tweet |
|---|---|
| **Topic 0, Theme: Views and Perspectives about MPox** | |
| Tweet #1 | @vancemurphy @pfizer @moderna_tx @US_FDA Well, you know the new thing is monkey pox, right? Vaccines are so yesterday. |
| Tweet #2 | Its annoys me how they use pictures of black peoples hands when they discuss monkey pox |
| Tweet #3 | The pics of monkey pox looks exactly like shingles |
| Tweet #4 | @masthahh1 Are there any stats on the people who have gotten monkey pox? Were they all vaccinated? |
| Tweet #5 | Looking at the state of the UK. I'd be more worried about Monkey Pox catching a dose of Englishman! |
| **Topic 1, Theme: Updates on Cases and Investigations about MPox** | |
| Tweet #1 | BREAKING: Health department investigating possible monkey pox case in NYC |
| Tweet #2 | New York health officials are investigating a potential case of monkeypox after a patient tested positive for the family of viruses associated with the rare illness. |
| Tweet #3 | U.S. government officials are placing orders for millions of doses of monkeypox vaccines amid a worldwide outbreak and a possible case in New York City, the Independent reports. |
| Tweet #4 | WHO is convening an Emergency Committee meeting out of concern for international spread of monkeypox, a high consequence infection. They will likely discuss whether to declare monkeypox a Public Health Emergency of International Concern (PHEIC) |
| Tweet #5 | The UK Health Security Agency said the new cases of the rare monkeypox infection do not have known connections with the previous confirmed cases announced on 14 May and a case on 7 May |
| **Topic 2, Theme: MPox and the LGBTQIA+ Community** | |
| Tweet #1 | @CraigbryCraig @BreezerGalway Moneypox has been known about since 1958. Majority of case are in gay males. No need to freak out |
| Tweet #2 | . Gay? Had "close" contact with someone whose in the hospital now in Montreal. Apparently majority in Montreal who contracted the Monkey Pox were gay 35–50 year old men. AIDS started in the gay community too. Something about monkeying around… |
| Tweet #3 | @jmcrookston Just to be SUPER CLEAR, what I mean by this, is that no, monkeypox isn't a "gay disease". I'm queer and super not okay with the way the media is framing this the same way HIV/AIDS was framed in the 70s/80s. |
| Tweet #4 | @jeffreyatucker @ezralevant Some knowledge about Monkey pox, it's mostly for gay. Not a threat. |
| Tweet #5 | @EnemyInAState @TimothyVollmer Absolutely agree only other events won't have the stigma attached which is happening with monkey pox so many people are convinced it's a gay disease because there's no context |
| **Topic 3, Theme: MPox and COVID-19** | |
| Tweet #1 | @COVIDnewsfast Transmission of Monkey Pox is not the same as Covid! |
| Tweet #2 | MUST WATCH: Amazing Polly exposes 2021 DAVOS pandemic event for a May 15 2022 release of MonkeyPox! BOOM! This Monkey Pox, like COVID, is being exploited to push a global government. Amazing Polly catches them. |
| Tweet #3 | @ANCParliament What are your plans on preventing Monkey Pox that has "accidentally" been released in the United States of America from coming into South Africa before it becomes a big Issue like Covid-19? |
| Tweet #4 | WW3 is on the horizon, Covid-19, and Monkey Pox about to be released into the world we'll be lucky if any humans survive? Nice work Joe. #LetsGoBrandon |
| Tweet #5 | Monkey Pox is coming! Covid did not do the trick. |

[1] These Tweets are presented here in "as is" form. These Tweets do not represent or reflect the views or opinions or beliefs or political stance of the author of this paper.

Figure 9 shows an analysis of the number of Tweets per topic. As can be seen from Figure 9, Topic 0, or the theme of *Views and Perspectives about Mpox*, was most popular on Twitter (in terms of the number of Tweets posted) during the time range of 7 May 2022 to 3 March 2023. It was followed by Topic 2, or the theme of *MPox and the LGBTQIA+ Community*. This was followed by Topic 3 (or the theme of *MPox and COVID-19*) and Topic 1 (or the theme of *Updates on Cases and Investigations about Mpox*), respectively.

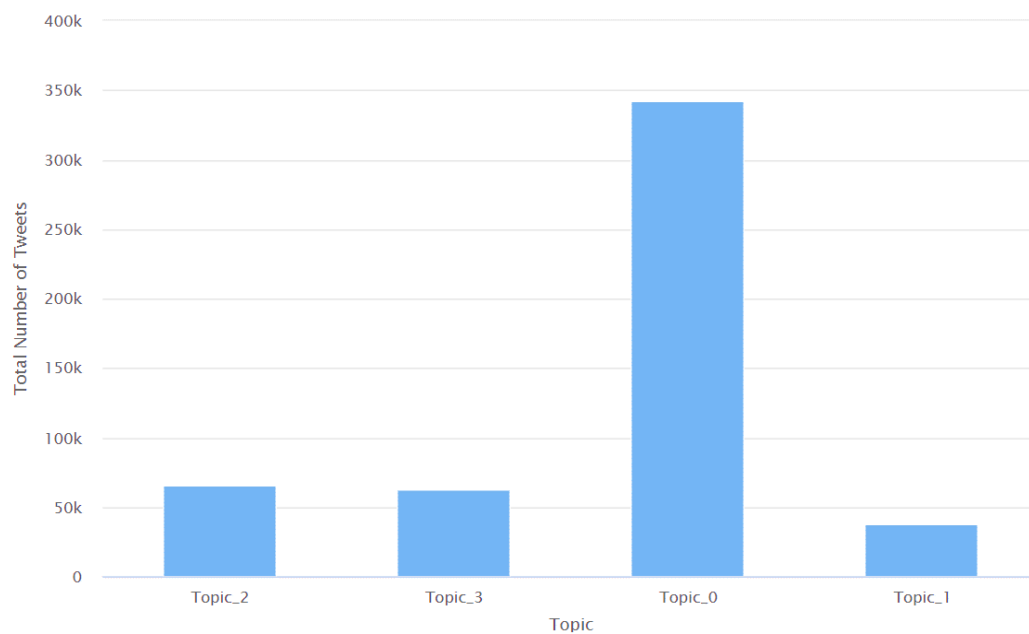

**Figure 9.** Analysis of the number of Tweets posted per topic.

Table 4 presents the characteristic features of this LDA model to outline its overall performance and working. The characteristics reported are average coherence scores, word lengths, exclusivity, document entropy, and tokens for each topic.

**Table 4.** Different characteristic features of the developed LDA model for each topic.

| Topic | Average Coherence | Average Word Length | Exclusivity | Document Entropy | Tokens |
| --- | --- | --- | --- | --- | --- |
| Topic 0 | −5.017 | 4 | 0.925 | 12.707 | 2,146,002 |
| Topic 1 | −2.000 | 7 | 0.972 | 11.084 | 310,322 |
| Topic 2 | −5.730 | 4 | 0.899 | 11.828 | 902,967 |
| Topic 3 | −5.946 | 4.667 | 0.855 | 11.758 | 572,057 |

Finally, a comparative study with related works in this area of research is presented. Table 5 outlines a summary of related studies in this area of research that focused on the mining and analysis of Tweets about MPox (reviewed in Section 2). As can be seen from Table 5, most of the recent studies in this area of research have focused on sentiment analysis and content analysis of Tweets. At the same time, a couple of works [109,113] also exist in this field where topic modeling of Tweets about MPox was performed. As can be seen from Table 5, topic modeling of Tweets about MPox is a research area that remains less explored and less investigated. A comparison of the different characteristics of this work and these two works [109,113] is presented in Table 6.

**Table 5.** Summary and categorization of the recent studies in this area of research.

| Work | Sentiment Analysis | Content Analysis | Topic Modeling | Dataset Development |
|---|---|---|---|---|
| Knudsen et al. [104] | | ✓ | | |
| Zuhanda et al. [105] | ✓ | | | |
| Ortiz-Martínez et al. [106] | | ✓ | | |
| Rahmanian et al. [107] | | ✓ | | |
| Cooper et al. [108] | ✓ | ✓ | | |
| Ng et al. [109] | | | ✓ | |
| Bengesi et al. [110] | ✓ | | | |
| Olusegun et al. [11] | ✓ | | | |
| Farahat et al. [112] | ✓ | ✓ | | |
| Sv et al. [113] | ✓ | | ✓ | |
| Mohney et al. [114] | ✓ | | | |
| Nia et al. [115] | | | | ✓ |
| Iparraguirre-Villanueva [116] | ✓ | | | |
| AL-Ahdal [117] | | ✓ | | |

**Table 6.** Comparison of specific characteristics of this work with two similar studies in this area of research that also focused on topic modeling of Tweets.

| Work | Number of Tweets Used for Topic Modeling | Time Range of the Tweets |
|---|---|---|
| Ng et al. [109] | 352,182 Tweets | 6 May 2022 to 23 July 2022 |
| Sv et al. [113] | 128,037 Tweets | 1 June 2022 to 25 June 2022 |
| Thakur et al. [this work] | 601,432 Tweets | 7 May 2022 to 3 March 2023 |

Table 6 outlines two limitations in a couple of similar works [109,113] in this field and highlights how the work of this paper addresses the same. The following is a comprehensive analysis of the limitations in these works [109,113]:

1. Limited time range of the analyzed Tweets: The time range of the Tweets that were analyzed in these works represents Tweets that were posted only during certain months of the 2022 MPox outbreak. One of the works [109] included Tweets that were posted on the day the first case of the 2022 MPox outbreak was recorded (7 May 2022) but the other work [113] did not. Furthermore, none of these works analyzed Tweets posted after 23 July 2022.

2. Limited number of Tweets used for topic modeling: The number of Tweets that were used for topic modeling in these works is 352,182 and 128,037 Tweets, respectively. It is relevant to mention here that the work presented in [113] involved a two-step process. In the first step, the authors analyzed a dataset of 556,402 Tweets about MPox and performed sentiment analysis of those Tweets. The results showed that 128,037 Tweets (23.01%) had a negative sentiment. Thereafter, only these 128,037 Tweets that had a negative sentiment were used for topic modeling in the second step of that study. The number of Tweets used for topic modeling in the previous works represents a fraction of the total number of Tweets that have been posted since the first recorded case of the 2022 Mpox outbreak on 7 May 2022.

3. Elimination of a topic from the study: The work of Ng et al. [109] reports that after performing topic modeling, one topic was categorized as "Miscellaneous", which accounted for 31.1% of the total number of Tweets. The work also reports that this topic was omitted from the results or, in other words, the specific themes or focus areas of conversation reported in that study [109] are based on the analysis of 68.9% of the Tweets only.

4. Lack of reporting of metrics to discuss the working or the accuracy of the topic modeling approaches: The average coherence value of an LDA model serves as a key indicator for the determination of the optimal number of topics. At the same time, metrics such as exclusivity, document entropy, number of tokens, and average word length of each topic help to provide a better understanding of the working of the underlying topic modeling approach. The two prior works that exist in this field [109,113] do not report any of these metrics to discuss either the working or the accuracy of the topic modeling approaches that were used.

These limitations that exist in similar studies in this area of research are addressed in this paper. First, the dataset that was used for developing the LDA model comprises Tweets about MPox that were posted on Twitter between 7 May 2022 and 3 March 2023—a time range that is greater than the time ranges of the similar works shown in Table 6. Second, a total of 601,432 Tweets were used for topic modeling. This is much higher than the number of Tweets that were used for topic modeling in similar works (352,182 Tweets in [109] and 128,037 Tweets in [113]) in this field. Third, no topic was eliminated or omitted prior to the identification of the themes or focus areas of conversations on Twitter related to MPox; in other words, the results presented in this work are based on the analysis of 100% of the Tweets present in the dataset. Finally, this work reports multiple metrics to discuss the accuracy and working of the topic modeling approach. Specifically, the LDA model developed in this paper was run by varying the number of topics from 2 to 50 and the average coherence score was computed for each topic. Based on the analysis of these coherence scores, the optimal number of topics was determined to be four. Thereafter, the LDA model was developed by considering the number of topics to be four and the average coherence scores, word lengths, exclusivity, document entropy, and tokens for each topic were reported to further discuss the accuracy and working of the developed LDA model. Thus, to summarize, the time range of the Tweets, the number of Tweets that were analyzed in this study, the fact that no topics were eliminated prior to the identification of themes of conversation, and the reporting of the average coherence scores, word lengths, exclusivity, document entropy, and tokens for each topic in the developed LDA model further support the scientific contributions of this work.

The discourse surrounding the global outbreak of MPox on the Twitter platform has garnered worldwide attention. The findings of this research reveal that conversations on Twitter revolving around MPox are characterized by their multifaceted nature, with the public actively engaged in seeking and disseminating information across a spectrum of topics connected to this virus. It is imperative for public health authorities to comprehend the primary concerns and interests of the populace concerning MPox. Such insights can serve as a foundation for the timely development of applicable policies and measures to bolster public awareness, preparedness, and response mechanisms in the face of the continued spread of the MPox virus and the potential resurgence thereof. Researchers from diverse domains have analyzed the mechanisms governing the flow of information from social media platforms to mainstream news outlets [132]. In today's contemporary media landscape, the once-distinct boundary separating social and mainstream media has blurred into obscurity. As presented in this study, one of the focal points of discussion on Twitter related to MPox pertains to updates on cases and investigations concerning the virus. This observation indicates that social media discourse mirrored news narratives (news regarding the latest developments and case reports of MPox) during this outbreak. Public health agencies could consider exploring this trend further by monitoring Twitter conversations, especially on days marked by significant news events, such as updates on vaccine developments or treatments or reports of severe adverse reactions attributed to specific vaccines or treatments for MPox. The identification of immediate reactions on Twitter may be helpful to these agencies for the development of timely responses and applicable policies. In view of the fact that the outbreak of MPox started a few months ago, the scientific community is still in the process of studying this virus. Consequently, there exists a spectrum of uncertainties encompassing the virus's transmission dynamics, current treatment modalities, vaccination strategies, and public receptiveness toward these treatments and vaccinations. These uncertainties have catalyzed an outpouring of opinions and perspectives on MPox, both directly concerning the virus and its outbreak. Public health authorities could also consider an exploration of Tweets related to these specific themes to gauge whether the ongoing public discourse serves to bridge the information gap between the public's needs and the information disseminated by various healthcare and medical sectors. Moreover, such an investigation could be instrumental in identifying instances of misinformation or the propagation of conspiracy theories in connection with the MPox outbreak.

Even though the limitations of the previous and related works in this area of research have been addressed in this paper, this work also has limitations. The Tweets analyzed in this paper were available on Twitter at the time of data analysis. However, Twitter allows users to delete their Tweets as well as to delete their accounts. Furthermore, as per Twitter's inactive account policy [133], accounts on Twitter that have been inactive for a very long time may be permanently removed, which results in the deletion of all the Tweets from that account. So, if this study is repeated in a few months or a few years from now, it is possible that the results obtained could vary to a degree if any of the analyzed Tweets were deleted due to the users of those accounts deleting those Tweets, users (who posted one or more of the analyzed Tweets) deleting their accounts, or Twitter permanently removing one or more of the accounts (from which one or more of the analyzed Tweets were posted) due to very long inactivity.

## 5. Conclusions

In the last decade and a half, the world has experienced outbreaks of a range of viruses, such as COVID-19, H1N1, flu, Ebola, Zika virus, Middle East Respiratory Syndrome (MERS), measles, and West Nile virus, just to name a few. In today's Internet of Everything era, the popularity of social media platforms has been growing exponentially. Social media platforms have served as virtual communities during the outbreaks of such viruses in the past, allowing people from different parts of the world to share and exchange information, news, perspectives, opinions, ideas, and comments related to the outbreaks. Researchers from different disciplines have analyzed this Big Data of conversations related to virus outbreaks on social media platforms such as Twitter, using concepts such as topic modeling to understand the underlying themes of conversations and information exchange that the general public participate in. The recent outbreak of the MPox virus has resulted in a tremendous increase in the utilization of social media platforms such as Twitter. Recent studies in this area of research have primarily focused on sentiment analysis and content analysis of Tweets about MPox, whereas a couple of studies in this area of research that have focused on topic modeling have multiple limitations. This paper aims to address this research gap and makes two scientific contributions to this field. First, it presents the results of performing topic modeling of 601,432 Tweets about the 2022 MPox outbreak that were posted on Twitter between 7 May 2022 and 3 March 2023. These results indicate that the conversations related to MPox during this time range may be broadly categorized into four distinct themes—*Views and Perspectives about MPox*, *Updates on Cases and Investigations about MPox*, *MPox and the LGBTQIA+ Community*, and *MPox and COVID-19*. Second, the paper presents the findings from the analysis of the Tweets that focused on these topics. The results show that the theme that was most popular on Twitter (in terms of the number of Tweets posted) during this time range was *Views and Perspectives about MPox*. It was followed by the theme of *MPox and the LGBTQIA+ Community*. This theme was followed by the themes of *MPox and COVID-19* and *Updates on Cases and Investigations about Mpox*, respectively. As per the best knowledge of the authors, no similar work has been conducted in this field thus far. With the continuous advances in the fields of healthcare and medicine in the last few months, the general public now has access to different forms of treatment and vaccines for MPox. As more treatments and vaccines become available to the public and studies reporting the accuracy and any potential side effects of the same are published, the patterns of conversations on Twitter related to MPox are expected to see an increase in the number of Tweets related to vaccines and treatments for MPox. Therefore, future work in this area would involve collecting more Tweets and repeating this study a few months from now to identify any significant variations in terms of the themes of conversations on Twitter related to MPox. If any new themes of conversations are identified, future work will also involve an exploration of the Tweets in those themes to understand the patterns of seeking and sharing information or misinformation related to those themes.

**Author Contributions:** Conceptualization, N.T.; methodology, N.T. and Y.N.D.; software, N.T.; validation, N.T.; formal analysis, N.T.; investigation, N.T.; resources, N.T. and Y.N.D.; data curation, N.T.; writing—original draft preparation, N.T., Y.N.D., and Z.L.; writing—review and editing, N.T., Y.N.D., and Z.L.; visualization, N.T.; supervision, N.T.; project administration, N.T.; funding acquisition, not applicable. All authors have read and agreed to the published version of the manuscript.

**Funding:** This research received no external funding.

**Data Availability Statement:** The data analyzed in this study are publicly available at https://dx.doi.org/10.21227/16ca-c879.

**Conflicts of Interest:** The authors declare no conflicts of interest.

## References

1. McCollum, A.M.; Damon, I.K. Human Monkeypox. *Clin. Infect. Dis.* **2014**, *58*, 260–267. https://doi.org/10.1093/cid/cit703.
2. Beer, E.M.; Rao, V.B. A Systematic Review of the Epidemiology of Human Monkeypox Outbreaks and Implications for Outbreak Strategy. PLoS Negl. *Trop. Dis.* **2019**, *13*, e0007791. https://doi.org/10.1371/journal.pntd.0007791.
3. Likos, A.M.; Sammons, S.A.; Olson, V.A.; Frace, A.M.; Li, Y.; Olsen-Rasmussen, M.; Davidson, W.; Galloway, R.; Khristova, M.L.; Reynolds, M.G.; et al. A Tale of Two Clades: Monkeypox Viruses. *J. Gen. Virol.* **2005**, *86*, 2661–2672. https://doi.org/10.1099/vir.0.81215-0.
4. Heymann, D.L.; Szczeniowski, M.; Esteves, K. Re-Emergence of Monkeypox in Africa: A Review of the Past Six Years. *Br. Med. Bull.* **1998**, *54*, 693–702. https://doi.org/10.1093/oxfordjournals.bmb.a011720.
5. Mandja, B.-A.M.; Brembilla, A.; Handschumacher, P.; Bompangue, D.; Gonzalez, J.-P.; Muyembe, J.-J.; Mauny, F. Temporal and Spatial Dynamics of Monkeypox in Democratic Republic of Congo, 2000–2015. *Ecohealth* **2019**, *16*, 476–487. https://doi.org/10.1007/s10393-019-01435-1.
6. Nguyen, P.-Y.; Ajisegiri, W.S.; Costantino, V.; Chughtai, A.A.; MacIntyre, C.R. Reemergence of Human Monkeypox and Declining Population Immunity in the Context of Urbanization, Nigeria, 2017–2020. *Emerg. Infect. Dis.* **2021**, *27*, 1007. https://doi.org/10.3201/eid2704.203569.
7. Yong, S.E.F.; Ng, O.T.; Ho, Z.J.M.; Mak, T.M.; Marimuthu, K.; Vasoo, S.; Yeo, T.W.; Ng, Y.K.; Cui, L.; Ferdous, Z.; et al. Imported Monkeypox, Singapore. *Emerg. Infect. Dis.* **2020**, *26*, 1826–1830. https://doi.org/10.3201/eid2608.191387.
8. Saxena, S.K.; Ansari, S.; Maurya, V.K.; Kumar, S.; Jain, A.; Paweska, J.T.; Tripathi, A.K.; Abdel-Moneim, A.S. Re-emerging Human Monkeypox: A Major Public-health Debacle. *J. Med. Virol.* 2023, 95. https://doi.org/10.1002/jmv.27902.
9. Kozlov, M. Monkeypox Declared a Global Emergency: Will It Help Contain the Outbreaks? *Nature* **2022**. https://doi.org/10.1038/d41586-022-02054-7.
10. Multi-Country Outbreak of Mpox, External Situation Report #22-11 May 2023. Available online: https://www.who.int/publications/m/item/multi-country-outbreak-of-mpox--external-situation-report--22---11-may-2023 (accessed on 31 August 2023).
11. Liu, L.; Xu, Z.; Fuhlbrigge, R.C.; Peña-Cruz, V.; Lieberman, J.; Kupper, T.S. Vaccinia Virus Induces Strong Immunoregulatory Cytokine Production in Healthy Human Epidermal Keratinocytes: A Novel Strategy for Immune Evasion. *J. Virol.* **2005**, *79*, 7363–7370. https://doi.org/10.1128/jvi.79.12.7363-7370.2005.
12. MacLeod, D.T.; Nakatsuji, T.; Wang, Z.; di Nardo, A.; Gallo, R.L. Vaccinia Virus Binds to the Scavenger Receptor MARCO on the Surface of Keratinocytes. *J. Invest. Dermatol.* **2015**, *135*, 142–150. https://doi.org/10.1038/jid.2014.330.
13. Vaccines. Available online: https://www.cdc.gov/smallpox/clinicians/vaccines.html (accessed on 31 August 2023).
14. Berhanu, A.; Prigge, J.T.; Silvera, P.M.; Honeychurch, K.M.; Hruby, D.E.; Grosenbach, D.W. Treatment with the Smallpox Antiviral Tecovirimat (ST-246) Alone or in Combination with ACAM2000 Vaccination Is Effective as a Postsymptomatic Therapy for Monkeypox Virus Infection. *Antimicrob. Agents Chemother.* **2015**, *59*, 4296–4300. https://doi.org/10.1128/aac.00208-15.
15. O'Shea, J.; Filardo, T.D.; Morris, S.B.; Weiser, J.; Petersen, B.; Brooks, J.T. Interim Guidance for Prevention and Treatment of Monkeypox in Persons with HIV Infection—United States, August 2022. MMWR Morb. *Mortal. Wkly. Rep.* **2022**, *71*, 1023–1028. https://doi.org/10.15585/mmwr.mm7132e4.
16. Piccolo, A.J.L.; Chan, J.; Cohen, G.M.; Mgbako, O.; Pitts, R.A.; Postelnicu, R.; Wallach, A.; Mukherjee, V. Critical Elements of an Mpox Vaccination Model at the Largest Public Health Hospital System in the United States. *Vaccines* **2023**, *11*, 1138. https://doi.org/10.3390/vaccines11071138.

17. CDC Detection & Transmission of Mpox Virus during the 2022 Clade IIb Out. Available online: https://www.cdc.gov/poxvirus/mpox/about/science-behind-transmission.html (accessed on 31 August 2023).
18. Mohanto, S.; Faiyazuddin, M.; Dilip Gholap, A.; Jogi, D.; Bhunia, A.; Subbaram, K.; Gulzar Ahmed, M.; Nag, S.; Shabib Akhtar, M.; Bonilla-Aldana, D.K.; et al. Addressing the Resurgence of Global Monkeypox (Mpox) through Advanced Drug Delivery Platforms. *Travel Med. Infect. Dis.* **2023**, *56*, 102636. https://doi.org/10.1016/j.tmaid.2023.102636.
19. Fifth Meeting of the International Health Regulations (2005) (IHR) Emergency Committee on the Multi-Country Outbreak of Mpox (Monkeypox). Available online: https://www.who.int/news/item/11-05-2023-fifth-meeting-of-the-international-health-regulations-(2005)-(ihr)-emergency-committee-on-the-multi-country-outbreak-of-monkeypox-(mpox) (accessed on 31 August 2023).
20. Miraz, M.H.; Ali, M.; Excell, P.S.; Picking, R. A Review on Internet of Things (IoT), Internet of Everything (IoE) and Internet of Nano Things (IoNT). In Proceedings of the 2015 Internet Technologies and Applications (ITA), IEEE; Wrexham, UK, 8–11 September 2015; pp. 219–224.
21. Twitter: Number of Users Worldwide 2024. Available online: https://www.statista.com/statistics/303681/twitter-users-worldwide/ (accessed on 31 August 2023).
22. Hutchinson, A. New Study Shows Twitter Is the Most Used Social Media Platform among Journalists. Available online: https://www.socialmediatoday.com/news/new-study-shows-twitter-is-the-most-used-social-media-platform-among-journa/626245/ (accessed on 31 August 2023).
23. Biggest Social Media Platforms 2023. Available online: https://www.statista.com/statistics/272014/global-social-networks-ranked-by-number-of-users/ (accessed on 31 August 2023).
24. Lin, Y. 10 Twitter Statistics Every Marketer Should Know in 2023 [Infographic]. Available online: https://www.oberlo.com/blog/twitter-statistics (accessed on 13 September 2023).
25. Martin, M. 29 Twitter Stats That Matter to Marketers in 2023. Available online: https://blog.hootsuite.com/twitter-statistics/ (accessed on 13 September 2023).
26. Twitter 'Lurkers' Follow—and Are Followed by—Fewer Accounts. Available online: https://www.pewresearch.org/short-reads/2022/03/16/5-facts-about-twitter-lurkers/ft_2022-03-16_twitterlurkers_03/ (accessed on 13 September 2023).
27. Thakur, N. Sentiment Analysis and Text Analysis of the Public Discourse on Twitter about COVID-19 and MPox. *Big Data Cogn. Comput.* **2023**, *7*, 116. https://doi.org/10.3390/bdcc7020116.
28. Blei, D.M. Probabilistic Topic Models. *Commun. ACM* **2012**, *55*, 77–84. https://doi.org/10.1145/2133806.2133826.
29. Gretarsson, B.; O'Donovan, J.; Bostandjiev, S.; Höllerer, T.; Asuncion, A.; Newman, D.; Smyth, P. TopicNets: Visual Analysis of Large Text Corpora with Topic Modeling. *ACM Trans. Intell. Syst. Technol.* **2012**, *3*, 1–26. https://doi.org/10.1145/2089094.2089099.
30. Sievert, C.; Shirley, K.E. LDAvis: A Method for Visualizing and Interpreting Topics Available online: https://aclanthology.org/W14-3110.pdf (accessed on 13 September 2023).
31. Panichella, A.; Dit, B.; Oliveto, R.; Di Penta, M.; Poshynanyk, D.; De Lucia, A. How to Effectively Use Topic Models for Software Engineering Tasks? An Approach Based on Genetic Algorithms. In Proceedings of the 2013 35th International Conference on Software Engineering (ICSE), IEEE; San Francisco, CA, USA, 18–26 May 2013; pp. 522–531.
32. Silva, C.C.; Galster, M.; Gilson, F. Topic Modeling in Software Engineering Research. *Empir. Softw. Eng.* **2021**, *26*, 120. https://doi.org/10.1007/s10664-021-10026-0.
33. Linton, M.; Teo, E.G.S.; Bommes, E.; Chen, C.Y.; Härdle, W.K. Dynamic Topic Modelling for Cryptocurrency Community Forums. In *Applied Quantitative Finance*; Springer: Berlin/Heidelberg, Germany, 2017; pp. 355–372. ISBN 9783662544853.
34. Schnoering, H. Short Text Topic Modeling: Application to Tweets about Bitcoin. *arXiv* **2022**.
35. Kang, H.-J.; Han, J.; Kwon, G.H. Determining the Intellectual Structure and Academic Trends of Smart Home Health Care Research: Coword and Topic Analyses. *J. Med. Internet Res.* **2021**, *23*, e19625. https://doi.org/10.2196/19625.
36. Thakur, N.; Han, C.Y. A Simplistic and Cost-Effective Design for Real-World Development of an Ambient Assisted Living System for Fall Detection and Indoor Localization: Proof-of-Concept. *Information* **2022**, *13*, 363. https://doi.org/10.3390/info13080363.
37. Huynh, T.; Fritz, M.; Schiele, B. Discovery of Activity Patterns Using Topic Models. In Proceedings of the 10th International Conference on Ubiquitous Computing, Seoul, Republic of Korea, 21–24 September 2008; ACM: New York, NY, USA, 2008.
38. Thakur, N.Y.; Han, C. Pervasive Activity Logging for Indoor Localization in Smart Homes. In Proceedings of the 2021 4th International Conference on Data Science and Information Technology, Shanghai China, 23–25 July 2021; ACM: New York, NY, USA, 2021.
39. Goudarzvand, S.; St. Sauver, J.; Mielke, M.M.; Takahashi, P.Y.; Lee, Y.; Sohn, S. Early Temporal Characteristics of Elderly Patient Cognitive Impairment in Electronic Health Records. *BMC Med. Inform. Decis. Mak.* **2019**, *19*, 149. https://doi.org/10.1186/s12911-019-0858-0.
40. Thakur, N.; Han, C.Y. A Multimodal Approach for Early Detection of Cognitive Impairment from Tweets. In *Human Interaction, Emerging Technologies and Future Systems V*; Springer: Cham, Switzerland, 2022; pp. 11–19. ISBN 9783030855390.
41. Yun, E. Review of Trends in Physics Education Research Using Topic Modeling. *J. Balt. Sci. Educ.* **2020**, *19*, 388–400.
42. Chen, Y.; Yu, B.; Zhang, X.; Yu, Y. Topic Modeling for Evaluating Students' Reflective Writing: A Case Study of Pre-Service Teachers' Journals. In Proceedings of the Sixth International Conference on Learning Analytics & Knowledge—LAK '16, Edinburgh, UK, 25–29 April 2016; ACM Press: New York, New York, USA, 2016.

43. Zhao, W.; Zou, W.; Chen, J.J. Topic Modeling for Cluster Analysis of Large Biological and Medical Datasets. *BMC Bioinform.* **2014**, *15*, S11. https://doi.org/10.1186/1471-2105-15-s11-s11.
44. Zheng, B.; McLean, D.C., Jr; Lu, X. Identifying Biological Concepts from a Protein-Related Corpus with a Probabilistic Topic Model. *BMC Bioinform.* **2006**, *7*, 58. https://doi.org/10.1186/1471-2105-7-58.
45. Porturas, T.; Taylor, R.A. Forty Years of Emergency Medicine Research: Uncovering Research Themes and Trends through Topic Modeling. *Am. J. Emerg. Med.* **2021**, *45*, 213–220. https://doi.org/10.1016/j.ajem.2020.08.036.
46. Yao, L.; Zhang, Y.; Wei, B.; Zhang, W.; Jin, Z. A Topic Modeling Approach for Traditional Chinese Medicine Prescriptions. *IEEE Trans. Knowl. Data Eng.* **2018**, *30*, 1007–1021. https://doi.org/10.1109/tkde.2017.2787158.
47. Firdaniza, F.; Ruchjana, B.; Chaerani, D.; Radianti, J. Information Diffusion Model in Twitter: A Systematic Literature Review. *Information* **2021**, *13*, 13. https://doi.org/10.3390/info13010013.
48. Bokaee Nezhad, Z.; Deihimi, M.A. Twitter Sentiment Analysis from Iran about COVID 19 Vaccine. *Diabetes Metab. Syndr.* **2022**, *16*, 102367. https://doi.org/10.1016/j.dsx.2021.102367.
49. Wang, Y.; Guo, J.; Yuan, C.; Li, B. Sentiment Analysis of Twitter Data. *Appl. Sci.* **2022**, *12*, 11775. https://doi.org/10.3390/app122211775.
50. Manias, G.; Mavrogiorgou, A.; Kiourtis, A.; Symvoulidis, C.; Kyriazis, D. Multilingual Text Categorization and Sentiment Analysis: A Comparative Analysis of the Utilization of Multilingual Approaches for Classifying Twitter Data. *Neural Comput. Appl.* **2023**, *35*, 21415–21431. https://doi.org/10.1007/s00521-023-08629-3.
51. Rodrigues, A.P.; Fernandes, R.; Aakash; Abhishek; Shetty, A.; Atul; Lakshmanna, K.; Shafi, R.M. Real-Time Twitter Spam Detection and Sentiment Analysis Using Machine Learning and Deep Learning Techniques. *Comput. Intell. Neurosci.* **2022**, *2022*, 5211949. https://doi.org/10.1155/2022/5211949.
52. Mao, H.; Shuai, X.; Kapadia, A. Loose Tweets: An Analysis of Privacy Leaks on Twitter. In Proceedings of the 10th annual ACM workshop on Privacy in the electronic society, Chicago, IL, USA, 17 October 2011; ACM: New York, NY, USA, 2011.
53. Mendoza, M.; Poblete, B.; Castillo, C. Twitter under Crisis: Can We Trust What We RT? In Proceedings of the First Workshop on Social Media Analytics, Washington, DC, USA, 25 July 2010; ACM: New York, NY, USA, 2010.
54. Zagheni, E.; Garimella, V.R.K.; Weber, I. Bogdan State Inferring International and Internal Migration Patterns from Twitter Data. In Proceedings of the 23rd International Conference on World Wide Web, Seoul, Republic of Korea, 7–11 April 2014; ACM: New York, NY, USA, 2014.
55. Ibrahim, R.; Elbagoury, A.; Kamel, M.S.; Karray, F. Tools and Approaches for Topic Detection from Twitter Streams: Survey. *Knowl. Inf. Syst.* **2018**, *54*, 511–539. https://doi.org/10.1007/s10115-017-1081-x.
56. Thakur, N. Social Media Mining and Analysis: A Brief Review of Recent Challenges. *Information* **2023**, *14*, 484. https://doi.org/10.3390/info14090484.
57. Abu Samah, K.A.F.; Amirah Misdan, N.F.; Hasrol Jono, M.N.H.; Riza, L.S. The Best Malaysian Airline Companies Visualization through Bilingual Twitter Sentiment Analysis: A Machine Learning Classification. JOIV Int. *J. Inform. Vis.* **2022**, *6*, 130. https://doi.org/10.30630/joiv.6.1.879.
58. Bodaghi, A.; Oliveira, J. The Theater of Fake News Spreading, Who Plays Which Role? A Study on Real Graphs of Spreading on Twitter. Expert Syst. *Appl.* **2022**, *189*, 116110. https://doi.org/10.1016/j.eswa.2021.116110.
59. Collins, S.; DeWitt, J. Words Matter: Presidents Obama and Trump, Twitter, and U.s. Soft Power. *World Aff.* **2023**, *186*, 530–571. https://doi.org/10.1177/00438200231161631.
60. Berrocal-Gonzalo, S.; Zamora-Martínez, P.; González-Neira, A. Politainment on Twitter: Engagement in the Spanish Legislative Elections of April 2019. *Media Commun.* **2023**, *11*, 163–175. https://doi.org/10.17645/mac.v11i2.6292.
61. Chang, R.-C.; Rao, A.; Zhong, Q.; Wojcieszak, M.; Lerman, K. #RoeOverturned: Twitter Dataset on the Abortion Rights Controversy. *Proc. Int. AAAI Conf. Web Soc. Media* **2023**, *17*, 997–1005. https://doi.org/10.1609/icwsm.v17i1.22207.
62. Peña-Fernández, S.; Larrondo-Ureta, A.; Morales-i-Gras, J. Feminism, gender identity and polarization in TikTok and Twitter. *Comunicar* **2023**, *31*, 49–60. https://doi.org/10.3916/c75-2023-04.
63. Goetz, S.J.; Heaton, C.; Imran, M.; Pan, Y.; Tian, Z.; Schmidt, C.; Qazi, U.; Ofli, F.; Mitra, P. Food Insufficiency and Twitter Emotions during a Pandemic. *Appl. Econ. Perspect. Policy* **2023**, *45*, 1189–1210. https://doi.org/10.1002/aepp.13258.
64. Tao, W.; Peng, Y. Differentiation and Unity: A Cross-Platform Comparison Analysis of Online Posts' Semantics of the Russian–Ukrainian War Based on Weibo and Twitter. *Commun. Public* **2023**, *8*, 105–124. https://doi.org/10.1177/20570473231165563.
65. Yavuz, G.R.; Kocak, M.E.; Ergun, G.; Alemdar, H.; Yalcin, H.; Incel, O.D.; Akarun, L.; Ersoy, C. A Smartphone Based Fall Detector with Online Location Support. Available online: http://sensorlab.cs.dartmouth.edu/phonesense/papers/Yavuz-fall.pdf (accessed on 13 September 2023).
66. Thakur, N.; Han, C.Y. An Approach for Detection of Walking Related Falls during Activities of Daily Living. In Proceedings of the 2020 International Conference on Big Data, Artificial Intelligence and Internet of Things Engineering (ICBAIE), Fuzhou, China, 12–14 June 2020; pp. 280–283.
67. Albín-Rodríguez, A.-P.; De-La-Fuente-Robles, Y.-M.; López-Ruiz, J.-L.; Verdejo-Espinosa, Á.; Espinilla Estévez, M. UJAmI Location: A Fuzzy Indoor Location System for the Elderly. *Int. J. Environ. Res. Public Health* **2021**, *18*, 8326. https://doi.org/10.3390/ijerph18168326.

68. Thakur, N.; Han, C.Y. A Context Driven Indoor Localization Framework for Assisted Living in Smart Homes. In *HCI International 2020—Late Breaking Papers: Universal Access and Inclusive Design*; Springer: Cham, Switzerland, 2020; pp. 387–400. ISBN 9783030601485.
69. Tamplain, P.M.; Fears, N.E.; Robinson, P.; Chatterjee, R.; Lichtenberg, G.; Miller, H.L. #DCD/Dyspraxia in Real Life: Twitter Users' Unprompted Expression of Experiences with Motor Differences. *J. Mot. Learn. Dev.* **2023**, *1*, 1–14. https://doi.org/10.1123/jmld.2023-0008.
70. Thakur, N.; Han, C.Y. An Ambient Intelligence-Based Human Behavior Monitoring Framework for Ubiquitous Environments. *Information* **2021**, *12*, 81. https://doi.org/10.3390/info12020081.
71. Song, B.-K.; Yang, B.-I. A Study on Cognitive Activity Programs for Elderly Person with Mild Cognitive Impairment. *JRTDD* **2023**, *6*, 13–17.
72. Thakur, N.; Han, C.Y. An Intelligent Ubiquitous Activity Aware Framework for Smart Home. In *Human Interaction, Emerging Technologies and Future Applications III.*; Springer: Cham, Switzerland, 2021; pp. 296–302. ISBN 9783030553067.
73. Lu, X.; Brelsford, C. Network Structure and Community Evolution on Twitter: Human Behavior Change in Response to the 2011 Japanese Earthquake and Tsunami. *Sci. Rep.* **2014**, *4*, 6773. https://doi.org/10.1038/srep06773.
74. Thakur, N.; Han, C.Y. A Framework for Prediction of Cramps during Activities of Daily Living in Elderly. In Proceedings of the 2020 International Conference on Big Data, Artificial Intelligence and Internet of Things Engineering (ICBAIE), Fuzhou, China, 12–14 June 2020; pp. 284–287.
75. Contreras-Somoza, L.M.; Irazoki, E.; Toribio-Guzmán, J.M.; de la Torre-Díez, I.; Diaz-Baquero, A.A.; Parra-Vidales, E.; Perea-Bartolomé, M.V.; Franco-Martín, M.Á. Usability and User Experience of Cognitive Intervention Technologies for Elderly People with MCI or Dementia: A Systematic Review. *Front. Psychol.* **2021**, *12*, 636116. https://doi.org/10.3389/fpsyg.2021.636116.
76. Thakur, N.; Han, C.Y. Methodology for Forecasting User Experience for Smart and Assisted Living in Affect Aware Systems. In Proceedings of the 8th International Conference on the Internet of Things, Santa Barbara, CA, USA, 15–18 October 2018; ACM: New York, NY, USA, 2018.
77. Carpenter, J.P.; Krutka, D.G. How and Why Educators Use Twitter: A Survey of the Field. *J. Res. Technol. Educ.* **2014**, *46*, 414–434. https://doi.org/10.1080/15391523.2014.925701.
78. Thakur, N. A Large-Scale Dataset of Twitter Chatter about Online Learning during the Current COVID-19 Omicron Wave. *Data* **2022**, *7*, 109. https://doi.org/10.3390/data7080109.
79. Sama, T.B.; Konttinen, I.; Hiilamo, H. Alcohol Industry Arguments for Liberalizing Alcohol Policy in Finland: Analysis of Twitter Data. *J. Stud. Alcohol Drugs* **2021**, *82*, 279–287. https://doi.org/10.15288/jsad.2021.82.279.
80. Curtis, B.; Giorgi, S.; Buffone, A.E.K.; Ungar, L.H.; Ashford, R.D.; Hemmons, J.; Summers, D.; Hamilton, C.; Schwartz, H.A. Can Twitter Be Used to Predict County Excessive Alcohol Consumption Rates? *PLoS ONE* **2018**, *13*, e0194290. https://doi.org/10.1371/journal.pone.0194290.
81. Kamiński, M.; Muth, A.; Bogdański, P. Smoking, Vaping, and Tobacco Industry during COVID-19 Pandemic: Twitter Data Analysis. *Cyberpsychol. Behav. Soc. Netw.* **2020**, *23*, 811–817. https://doi.org/10.1089/cyber.2020.0384.
82. Myslín, M.; Zhu, S.-H.; Chapman, W.; Conway, M. Using Twitter to Examine Smoking Behavior and Perceptions of Emerging Tobacco Products. *J. Med. Internet Res.* **2013**, *15*, e174. https://doi.org/10.2196/jmir.2534.
83. Williams, J.; Chinn, S.J.; Suleiman, J. The Value of Twitter for Sports Fans. *J. Direct Data Digit. Mark. Pract.* **2014**, *16*, 36–50. https://doi.org/10.1057/dddmp.2014.36.
84. Hutchins, B. Twitter: Follow the Money and Look beyond Sports. *Commun. Sport* **2014**, *2*, 122–126. https://doi.org/10.1177/2167479514527430.
85. López-de-Ipiña, D.; Díaz-de-Sarralde, I.; García-Zubia, J. An Ambient Assisted Living Platform Integrating RFID Data-on-Tag Care Annotations and Twitter. Available online: https://citeseerx.ist.psu.edu/document?repid=rep1&type=pdf&doi=3484a2c17c27695659a925cb4892dc9d7ac318ec (accessed on 13 September 2023).
86. Thakur, N.; Han, C.Y. A Study of Fall Detection in Assisted Living: Identifying and Improving the Optimal Machine Learning Method. *J. Sens. Actuator Netw.* **2021**, *10*, 39. https://doi.org/10.3390/jsan10030039.
87. Aresta, M.; Pedro, L.; Santos, C.; Moreira, A. Portraying the Self in Online Contexts: Context-Driven and User-Driven Online Identity Profiles. *Contemp. Soc. Sci.* **2015**, *10*, 70–85. https://doi.org/10.1080/21582041.2014.980840.
88. Thakur, N.; Han, C.Y. A Multilayered Contextually Intelligent Activity Recognition Framework for Smart Home. In *Human Interaction, Emerging Technologies and Future Applications III.*; Springer: Cham, Switzerland, 2021; pp. 278–283. ISBN 9783030553067.
89. Weeks, R.; White, S.; Hartner, A.-M.; Littlepage, S.; Wolf, J.; Masten, K.; Tingey, L. COVID-19 Messaging on Social Media for American Indian and Alaska Native Communities: Thematic Analysis of Audience Reach and Web Behavior. *JMIR Infodemiology* **2022**, *2*, e38441. https://doi.org/10.2196/38441.
90. Thakur, N.; Han, C.Y. Country-Specific Interests towards Fall Detection from 2004–2021: An Open Access Dataset and Research Questions. *Data* **2021**, *6*, 92. https://doi.org/10.3390/data6080092.
91. Eriksson, H.; Salzmann-Erikson, M. Twitter Discussions about the Predicaments of Robots in Geriatric Nursing: Forecast of Nursing Robotics in Aged Care. *Contemp. Nurse* **2018**, *54*, 97–107. https://doi.org/10.1080/10376178.2017.1364972.
92. Thakur, N.; Han, C.Y. An Approach to Analyze the Social Acceptance of Virtual Assistants by Elderly People. In Proceedings of the 8th International Conference on the Internet of Things; ACM: New York, NY, USA, 2018.

93. Cevik, F.; Kilimci, Z.H. Analysis of Parkinson's Disease Using Deep Learning and Word Embedding Models. *Acad. Perspect. Procedia* **2019**, *2*, 786–797.
94. Kesler, S.R.; Henneghan, A.M.; Thurman, W.; Rao, V. Identifying Themes for Assessing Cancer-Related Cognitive Impairment: Topic Modeling and Qualitative Content Analysis of Public Online Comments. *JMIR Cancer* **2022**, *8*, e34828. https://doi.org/10.2196/34828.
95. Klein, A.Z.; Kunatharaju, S.; O'Connor, K.; Gonzalez-Hernandez, G. Pregex: Rule-Based Detection and Extraction of Twitter Data in Pregnancy. *J. Med. Internet Res.* **2023**, *25*, e40569. https://doi.org/10.2196/40569.
96. Thakur, N.; Han, C.Y. A Human-Human Interaction-Driven Framework to Address Societal Issues. In *Human Interaction, Emerging Technologies and Future Systems V*; Springer International Publishing: Cham, 2022; pp. 563–571 ISBN 9783030855390.
97. Thakur, N.; Han, C.Y. A Framework for Facilitating Human-Human Interactions to Mitigate Loneliness in Elderly. In *Human Interaction, Emerging Technologies and Future Applications III*; Springer: Cham, Switzerland, 2021; pp. 322–327. ISBN 9783030553067.
98. Klein, A.Z.; O'Connor, K.; Levine, L.D.; Gonzalez-Hernandez, G. Using Twitter Data for Cohort Studies of Drug Safety in Pregnancy: Proof-of-Concept with β-Blockers. *JMIR Form. Res.* **2022**, *6*, e36771. https://doi.org/10.2196/36771.
99. Gomide, J.; Veloso, A.; Meira, W., Jr; Almeida, V.; Benevenuto, F.; Ferraz, F.; Teixeira, M. Dengue Surveillance Based on a Computational Model of Spatio-Temporal Locality of Twitter. In Proceedings of the 3rd International Web Science Conference, Koblenz, Germany, 15–17 June 2011; ACM: New York, NY, USA, 2011.
100. Radzikowski, J.; Stefanidis, A.; Jacobsen, K.H.; Croitoru, A.; Crooks, A.; Delamater, P.L. The Measles Vaccination Narrative in Twitter: A Quantitative Analysis. *JMIR Public Health Surveill.* **2016**, *2*, e1. https://doi.org/10.2196/publichealth.5059.
101. Signorini, A.; Segre, A.M.; Polgreen, P.M. The Use of Twitter to Track Levels of Disease Activity and Public Concern in the U.S. during the Influenza A H1N1 Pandemic. *PLoS ONE* **2011**, *6*, e19467. https://doi.org/10.1371/journal.pone.0019467.
102. Sugumaran, R.; Voss, J. Real-Time Spatio-Temporal Analysis of West Nile Virus Using Twitter Data. In Proceedings of the 3rd International Conference on Computing for Geospatial Research and Applications, Washington, DC, USA, 1–3 July 2012; ACM: New York, NY, USA, 2012.
103. Porat, T.; Garaizar, P.; Ferrero, M.; Jones, H.; Ashworth, M.; Vadillo, M.A. Content and Source Analysis of Popular Tweets Following a Recent Case of Diphtheria in Spain. *Eur. J. Public Health* **2019**, *29*, 117–122. https://doi.org/10.1093/eurpub/cky144.
104. Knudsen, B.; Høeg, T.B.; Prasad, V. Analysis of Tweets Discussing the Risk of Mpox among Children and Young People in School (May–October 2022): Public Health Experts on Twitter Consistently Exaggerated Risks and Infrequently Reported Accurate Information. *bioRxiv* **2023**.
105. Zuhanda, M.K. Analysis of Twitter User Sentiment on the Monkeypox Virus Issue Using the Nrc Lexicon Available online: https://www.iocscience.org/ejournal/index.php/mantik/article/view/3502 (accessed on 31 August 2023).
106. Ortiz-Martínez, Y.; Sarmiento, J.; Bonilla-Aldana, D.K.; Rodríguez-Morales, A.J. Monkeypox Goes Viral: Measuring the Misinformation Outbreak on Twitter. *J. Infect. Dev. Ctries.* **2022**, *16*, 1218–1220. https://doi.org/10.3855/jidc.16907.
107. Rahmanian, V.; Jahanbin, K.; Jokar, M. Using Twitter and Web News Mining to Predict the Monkeypox Outbreak. *Asian Pac. J. Trop. Med.* **2022**, *15*, 236. https://doi.org/10.4103/1995-7645.346083.
108. Cooper, L.N.; Radunsky, A.P.; Hanna, J.J.; Most, Z.M.; Perl, T.M.; Lehmann, C.U.; Medford, R.J. Analyzing an Emerging Pandemic on Twitter: Monkeypox. *Open Forum Infect. Dis.* **2023**, 10. https://doi.org/10.1093/ofid/ofad142.
109. Ng, Q.X.; Yau, C.E.; Lim, Y.L.; Wong, L.K.T.; Liew, T.M. Public Sentiment on the Global Outbreak of Monkeypox: An Unsupervised Machine Learning Analysis of 352,182 Twitter Posts. *Public Health* **2022**, *213*, 1–4. https://doi.org/10.1016/j.puhe.2022.09.008.
110. Bengesi, S.; Oladunni, T.; Olusegun, R.; Audu, H. A Machine Learning-Sentiment Analysis on Monkeypox Outbreak: An Extensive Dataset to Show the Polarity of Public Opinion from Twitter Tweets. *IEEE Access* **2023**, *11*, 11811–11826. https://doi.org/10.1109/access.2023.3242290.
111. Olusegun, R.; Oladunni, T.; Audu, H.; Houkpati, Y.A.O.; Bengesi, S. Text Mining and Emotion Classification on Monkeypox Twitter Dataset: A Deep Learning-Natural Language Processing (NLP) Approach. *IEEE Access* **2023**, *11*, 49882–49894. https://doi.org/10.1109/access.2023.3277868.
112. Farahat, R.A.; Yassin, M.A.; Al-Tawfiq, J.A.; Bejan, C.A.; Abdelazeem, B. Public Perspectives of Monkeypox in Twitter: A Social Media Analysis Using Machine Learning. *New Microbes New Infect.* 2022, 49–50, 101053. https://doi.org/10.1016/j.nmni.2022.101053.
113. Sv, P.; Ittamalla, R. What Concerns the General Public the Most about Monkeypox Virus?—A Text Analytics Study Based on Natural Language Processing (NLP). *Travel Med. Infect. Dis.* **2022**, *49*, 102404. https://doi.org/10.1016/j.tmaid.2022.102404.
114. Mohbey, K.K.; Meena, G.; Kumar, S.; Lokesh, K. A CNN-LSTM-Based Hybrid Deep Learning Approach to Detect Sentiment Polarities on Monkeypox Tweets. *arXiv* **2022**.
115. Nia, Z.M.; Bragazzi, N.L.; Wu, J.; Kong, J.D. A Twitter Dataset for Monkeypox, May 2022. *Data Brief* **2023**, *48*, 109118. https://doi.org/10.1016/j.dib.2023.109118.
116. Iparraguirre-Villanueva, O.; Alvarez-Risco, A.; Herrera Salazar, J.L.; Beltozar-Clemente, S.; Zapata-Paulini, J.; Yáñez, J.A.; Cabanillas-Carbonell, M. The Public Health Contribution of Sentiment Analysis of Monkeypox Tweets to Detect Polarities Using the CNN-LSTM Model. *Vaccines* **2023**, *11*, 312. https://doi.org/10.3390/vaccines11020312.

117. AL-Ahdal, T.; Coker, D.; Awad, H.; Reda, A.; Żuratyński, P.; Khailaie, S. Improving Public Health Policy by Comparing the Public Response during the Start of COVID-19 and Monkeypox on Twitter in Germany: A Mixed Methods Study. *Vaccines* **2022**, *10*, 1985. https://doi.org/10.3390/vaccines10121985.
118. Mierswa, I.; Wurst, M.; Klinkenberg, R.; Scholz, M.; Euler, T. YALE: Rapid Prototyping for Complex Data Mining Tasks. In Proceedings of the 12th ACM SIGKDD international conference on Knowledge discovery and data mining, Philadelphia, PA USA, 20–23 August 2006; ACM: New York, NY, USA, 2006.
119. Thakur, N.; Han, C.Y. Multimodal Approaches for Indoor Localization for Ambient Assisted Living in Smart Homes. *Information* **2021**, *12*, 114. https://doi.org/10.3390/info12030114.
120. Garner, S.R. WEKA: The Waikato Environment for Knowledge Analysis. Available online: https://citeseerx.ist.psu.edu/document?repid=rep1&type=pdf&doi=b732d47f60ff8b3e0dbe09dd098578fb00a971f4 (accessed on 13 September 2023).
121. Kohavi, R.; Sommerfield, D. D2.1.2 MLC ++. Available online: https://citeseerx.ist.psu.edu/document?repid=rep1&type=pdf&doi=7fd85a0b6ab6b37dd7a940e6b2813917493cb7fe (accessed on 13 September 2023).
122. Jelodar, H.; Wang, Y.; Yuan, C.; Feng, X.; Jiang, X.; Li, Y.; Zhao, L. Latent Dirichlet Allocation (LDA) and Topic Modeling: Models, Applications, a Survey. *Multimed. Tools Appl.* **2019**, *78*, 15169–15211. https://doi.org/10.1007/s11042-018-6894-4.
123. Wei, X.; Croft, W.B. LDA-Based Document Models for Ad-Hoc Retrieval. In Proceedings of the 29th Annual International ACM SIGIR Conference on Research and Development in Information Retrieval, Seattle, DC, USA, 6–11 August 2006; ACM: New York, NY, USA, 2006.
124. Yao, L.; Mimno, D.; McCallum, A. Efficient Methods for Topic Model Inference on Streaming Document Collections. In Proceedings of the 15th ACM SIGKDD International Conference on Knowledge Discovery and Data Mining, Paris France, 28 June 2009–1 July 2009; ACM: New York, NY, USA, 2009.
125. Thakur, N. MonkeyPox2022Tweets: A Large-Scale Twitter Dataset on the 2022 Monkeypox Outbreak, Findings from Analysis of Tweets, and Open Research Questions. *Infect. Dis. Rep.* **2022**, *14*, 855–883. https://doi.org/10.3390/idr14060087.
126. Anupriya, P.; Karpagavalli, S. LDA Based Topic Modeling of Journal Abstracts. In Proceedings of the 2015 International Conference on Advanced Computing and Communication Systems, Washington, DC, USA, 26–27 February 2015; pp. 1–5.
127. Guo, H.; Liang, Q.; Li, Z. An Improved AD-LDA Topic Model Based on Weighted Gibbs Sampling. In Proceedings of the 2016 IEEE Advanced Information Management, Communicates, Electronic and Automation Control Conference (IMCEC), Xi'an, China, 3–5 October 2016; pp. 1978–1982.
128. Syed, S.; Spruit, M. Full-Text or Abstract? Examining Topic Coherence Scores Using Latent Dirichlet Allocation. In Proceedings of the 2017 IEEE International Conference on Data Science and Advanced Analytics (DSAA), Tokyo, Japan, 19–21 October 2017; pp. 165–174.
129. Omar, M.; On, B.-W.; Lee, I.; Choi, G.S. LDA Topics: Representation and Evaluation. *J. Inf. Sci.* **2015**, *41*, 662–675. https://doi.org/10.1177/0165551515587839.
130. Stevens, K.; Kegelmeyer, P.; Andrzejewski, D.; Buttler, D. Exploring Topic Coherence over Many Models and Many Topics. Available online: https://aclanthology.org/D12-1087.pdf (accessed on 14 September 2023).
131. Xue, J.; Chen, J.; Chen, C.; Zheng, C.; Li, S.; Zhu, T. Public Discourse and Sentiment during the COVID 19 Pandemic: Using Latent Dirichlet Allocation for Topic Modeling on Twitter. *PLoS ONE* **2020**, *15*, e0239441. https://doi.org/10.1371/journal.pone.0239441.
132. Orellana-Rodriguez, C.; Keane, M.T. Attention to News and Its Dissemination on Twitter: A Survey. *Comput. Sci. Rev.* **2018**, *29*, 74–94. https://doi.org/10.1016/j.cosrev.2018.07.001.
133. Twitter's Inactive Account Policy. Available online: https://help.twitter.com/en/rules-and-policies/inactive-x-accounts (accessed on 13 September 2023).